%% file: main.tex
\documentclass[onefignum,onetabnum]{siamonline190516}

\input{shared}

\ifpdf
\hypersetup{
  pdftitle={Moments of Uniform Random Multigraphs with Fixed Degree Sequences},
  pdfauthor={P. S. Chodrow}
}
\fi

\myexternaldocument{supplement}

\begin{document}

\maketitle

\begin{abstract}
    We study the expected adjacency matrix of a uniformly random multigraph with fixed degree sequence  $\degree \in \Z_+^n$. 
    This matrix arises in a variety of analyses of networked data sets, including modularity-maximization and mean-field theories of spreading processes.  
    Its structure is well-understood for large, sparse, simple graphs: the expected number of edges between nodes $i$ and $j$ is roughly $\frac{d_id_j}{\sum_\ell{d_\ell}}$.
    Many network data sets are neither large, sparse, nor simple, and in these cases the standard approximation no longer applies. 
    We derive a novel estimator using a dynamical approach: the estimator emerges from the stationarity conditions of a class of Markov Chain Monte Carlo algorithms for graph sampling. 
    We derive error bounds for this estimator, and provide an efficient scheme with which to compute it. 
    We test the estimator on synthetic and empirical degree sequences, finding that it enjoys relative error against ground truth a full order of magnitude smaller than the standard approximation.
    We then compare modularity maximization techniques using both the standard and novel estimator, finding that the qualitative structure of the optimization landscape depends significantly on the estimator choice.   
    Our results emphasize the importance of using carefully specified random graph models in data scientific applications.
\end{abstract}



\subfile{tex/intro}

\subfile{tex/random_graphs}

\subfile{tex/error_bounds}

\subfile{tex/algorithms}

\subfile{tex/results}

\subfile{tex/discussion}

    \section*{Software and Data}
    
        We used the implementation of MCMC in \cite{Chodrow2019} to conduct simulation experiments. 
        All additional code used in this study may be freely accessed at \url{https://github.com/PhilChodrow/multigraph_moments}.
        The data used in this study is freely available courtesy of the authors of \cite{Benson2018} at \url{https://www.cs.cornell.edu/~arb/data/index.html}. 

\nocite{bhatia2013matrix,horn2012matrix}    

\bibliographystyle{siamplain}
\bibliography{references}

\appendix

\section{Proof of \Cref{thm:W}} \label{subsec:thm_W_proof}

      Write 
      \begin{align}
        \E[w_{ij}\Delta_{ij}] &= \E[w_{ij}(\Delta_{ij}^+ - \Delta_{ij}^-)]\\
         &= \frac{1}{z(\W)}\E\left[w_{ij}\left(b_ib_j  - 2y  + b_i + b_j - \x_i^T\x_j - 2\right)\right]\;. \label{eq:EWD}   
      \end{align}
      We have used the identity $w_{ij}x_{ij} =  w_{ij}$ to eliminate instances of $x_{ij}$ from this expression. 

      Starting with the first term, write 
      \begin{align*}
        \E[w_{ij}b_ib_j] &= \omega_{ij}\E[b_ib_j]\left(1 + \frac{\cov{w_{ij},b_ib_j}}{\omega_{ij}\E[b_ib_j]}\right) \\ 
        &= \omega_{ij}\beta_i\beta_j \left(\frac{\E[b_ib_j]}{\beta_i\beta_j} + \frac{\cov{w_{ij},b_ib_j}}{\omega_{ij}\beta_i\beta_j} \right)\;. 
      \end{align*}
      Applying \eqref{eq:ij}, the first term inside  the parentheses satisfies
      \begin{align*}
        \abs{\frac{\E[b_ib_j]}{\beta_i\beta_j} - 1} \leq \frac{2u_*}{\beta_i\vee\beta_j}\;.
      \end{align*}
      To bound the second term, we apply  Cauchy-Schwartz along with \eqref{eq:ij} and \eqref{eq:iijj}. 
      \begin{align*}
        \abs{\cov{w_{ij},b_ib_j}} &\leq \sigma_{ij}\sqrt{\var{b_ib_j}} \\ 
        &\leq \sqrt{(\beta_i + 6u_*)^2(\beta_j+6u_*)^2 - (\beta_i\beta_j - 2u_*(\beta_i\wedge\beta_j))^2}\;.
      \end{align*}
      We therefore obtain 
      \begin{align*}
        \abs{\E[w_{ij}b_ib_j] - \omega_{ij}\beta_i\beta_j} \leq \frac{2u_*}{\beta_i\vee\beta_j} + \frac{\sigma_{ij}}{\omega_{ij}}\frac{\sqrt{(\beta_i + 6u_*)^2(\beta_j+6u_*)^2 - (\beta_i\beta_j - 2u_*(\beta_i\wedge\beta_j))^2}}{\beta_i\beta_j}\;.
      \end{align*}

      The remaining terms are simpler. 
      We have 
      \begin{align*}
        \abs{\E[w_{ij}y] - \omega_{ij}\psi} &= \abs{\cov{w_{ij},y}}\leq \sigma_{ij}\sqrt{\var{y}} \leq \sigma_{ij}\sqrt{2v_*\psi}\;,
      \end{align*}
      where in the first inequality we have used Cauchy-Schwartz and in the second we have used \eqref{eq:var_y}.

      To control the remaining terms, start by using the same argument to yield 
      \begin{align*}
        \abs{\E[w_{ij}b_i] - \omega_{ij}\beta_i} \leq \sigma_{ij} \sqrt{2u_*\beta_i}\;. 
      \end{align*}
      Then, we obtain the upper bound
      \begin{align*}
        \E[w_{ij}(b_i + b_j) - \x_i^T\x_j - 2] &\leq \E[w_{ij}(b_i + b_j)] \\ 
        &\leq \omega_{ij}(\beta_i + \beta_j) + \sigma_{ij}\sqrt{2u_*}(\sqrt{\beta_i} + \sqrt{\beta_j})\;.
      \end{align*}
      Inserting our results into \eqref{eq:EWD}, we are able to write 
      \begin{align}
        \abs{\E[w_{ij}\Delta_{ij}]- \frac{\omega_{ij}(\beta_i\beta_j - 2\psi)}{z(\W)}} &\leq \frac{\epsilon'_{ij}(\BETA)}{z(\W)}\;, \label{eq:intermediate}
      \end{align}
      where 
      \begin{align*}
        \epsilon'_{ij}(\BETA) &\triangleq \frac{2u_*}{\beta_i\vee\beta_j} + \frac{\sigma_{ij}}{\omega_{ij}}\frac{\sqrt{(\beta_i + 6u_*)^2(\beta_j+6u_*)^2 - (\beta_i\beta_j - 2u_*(\beta_i\wedge\beta_j))^2}}{\beta_i\beta_j} \\ 
        &\quad  + \sigma_{ij}\sqrt{2v_*\psi_i} + \omega_{ij}(\beta_i + \beta_j) + \sigma_{ij}\sqrt{2u_*}(\sqrt{\beta_i} + \sqrt{\beta_j})]\;.
      \end{align*}
      From \Cref{lm:delta}, we also have 
      \begin{align*}
        \abs{\delta_{ij}^+ + \delta_{ij}^- - \frac{\beta_i\beta_j + 2\psi\chi_{ij}}{z(\W)}} \leq \frac{\epsilon_{ij}^-(\BETA) + \epsilon_{ij}^+(\BETA)}{z(\W)}\;.
      \end{align*}
      By definition, the righthand side is $\epsilon_{ij}^{\BETA}/z(\W)$. 
      Inserting these results into the stationarity condition \eqref{eq:p=2} and combining with \eqref{eq:intermediate}, we obtain 
      \begin{align*}
        \abs{\frac{2\omega_{ij}(\beta_i\beta_j - 2\psi) - \beta_i\beta_j - 2\psi\chi_{ij}}{z(\W)}} \leq \frac{2\epsilon_{ij}'(\BETA) + \epsilon_{ij}(\BETA)}{z(\W)}\;.
      \end{align*} 
      Dividing through by $2z(\W)^{-1}\abs{\beta_i\beta_j - 2\psi}$ and rearranging yields 
      \begin{align*}
        \abs{\omega_{ij} - \frac{1}{2}\frac{f_{ij}(\BETA) + \chi_{ij}}{1 - f_{ij}(\BETA)}} \leq \frac{2\epsilon'_{ij}(\BETA) + \epsilon_{ij}(\BETA)}{2\psi(1-f_{ij}(\BETA))}\;.
      \end{align*}
      Finally, we apply \Cref{thm:X} to approximate $\chi_{ij}$, 
      obtaining 
      \begin{align*}
        \abs{\omega_{ij} - \frac{f_{ij}(\BETA)}{1 - f_{ij}(\BETA)}} &\leq \frac{2\epsilon'_{ij}(\BETA) + \epsilon_{ij}(\BETA)}{2\psi(1-f_{ij}(\BETA))} + \frac{1}{2}\frac{\epsilon_{ij}(\BETA)}{1-f_{ij}(\BETA)} \\ 
        &= \frac{1}{1-f_{ij}(\BETA)} \left(\frac{2\epsilon'_{ij}(\BETA) + \epsilon_{ij}(\BETA)}{2\psi} + \frac{\epsilon_{ij}(\BETA)}{2}\right)\;,
      \end{align*}
      completing the proof. 

\section{Proof of \Cref{lm:jacobian}}\label{sec:SI_jacobian}
  We will first derive the expression for the Jacobian of $\mathbf{h}$ given in the text as \eqref{eq:deriv}.  
  For notational compactness, let $x_{ij} = f_{ij}(\BETA)$. 
  We first calculate 
  \begin{align*}
      \frac{\partial x_{ij}}{\partial \beta_k} = 
    \begin{cases}
      -\frac{x_{ij}}{2\psi} &\quad k \neq i,j\\ 
      x_{ij}(\frac{1}{\beta_i} -\frac{1}{2\psi}) &\quad k = i\\ 
      x_{ij}(\frac{1}{\beta_j} -\frac{1}{2\psi}) &\quad k = j\;. 
    \end{cases}
  \end{align*}
  Next, 
  \begin{align*}
    \frac{\partial}{\partial \beta_k}\left[\frac{x_{ij}}{1-x_{ij}}\right] = \frac{1}{(1-x_{ij})^2}\frac{\partial x_{ij}}{\partial \beta_k}\;.
  \end{align*}
  We can therefore write the components of the Jacobian as 
  \begin{align*}
    \frac{\partial h_i(\BETA)}{\partial \beta_k} = 
    \begin{cases}
      \left(\frac{1}{\beta_i} - \frac{1}{2\psi}\right)\sum_{j \neq i} \frac{x_{ij}}{(1-x_{ij})^2}
        &\quad k = i \\ 
        \frac{1}{\beta_k}\frac{f_{ik}}{(1-f_{ik})^2} - \frac{1}{2\psi}\sum_{j \neq i} \frac{x_{ij}}{(1-x_{ij})^2}&\quad k \neq i\;.
    \end{cases}
  \end{align*}
  It is convenient to define the matrix $\mathbf{S}\in \R^{n\times n}$ with entries $s_{ij} = \frac{x_{ij}}{(1-x_{ij})^2}$, in which case we can write 
  \begin{align*}
    \frac{\partial h_i(\BETA)}{\partial \beta_k} = \begin{cases}
      \left(\frac{1}{\beta_i} - \frac{1}{2\psi}\right)\sum_{j \neq i} s_{ij} &\quad k = i \\ 
        \frac{1}{\beta_k}a_{ik} - \frac{1}{2\psi}\sum_{j \neq i} s_{ij}&\quad k \neq i\;.
    \end{cases}
  \end{align*}
  Let $\mathbf{B} = \diag \BETA$ and $\mathbf{D} = \diag \mathbf{A}\mathbf{e}$. 
  We can then write the Jacobian in matrix form as 
  \begin{align*}
    \mathbf{J} =  (\mathbf{S} + \mathbf{D})\mathbf{B}^{-1} - \frac{1}{2\psi}\mathbf{S}\mathbf{E}\;.
  \end{align*}
  We next note that $\mathbf{S}\mathbf{E} = \mathbf{D}\mathbf{E}$, and therefore substitute $\mathbf{S}\mathbf{E} = \frac{1}{2}(\mathbf{S}\mathbf{E} + \mathbf{D}\mathbf{E})$. 
  The result is 
  \begin{align*}
    \mathbf{J} =  (\mathbf{S} + \mathbf{D})\left(\mathbf{B}^{-1} - \frac{1}{4\psi}\mathbf{E}\right)\;,
  \end{align*}
  as was to be shown.

  We now proceed with the proof of \Cref{lm:jacobian}. 
  We will employ the crude lower bound (see \cite{bhatia2013matrix}):
  \begin{align*}
    \lambda_1(\mathbf{J}) \geq \lambda_1\left(\mathbf{A} + \mathbf{D}\right)\lambda_1\left(\mathbf{B}^{-1} - \frac{1}{4\psi}\mathbf{E}\right)\;.
  \end{align*}
  Starting with the second factor, we can obtain an analytical inverse  (using, for example, the Sherman-Morrison formula): 
  \begin{align*}
    \left(\mathbf{B}^{-1} - \frac{1}{4\psi}\mathbf{E}\right)^{-1} =\mathbf{B} + \frac{\mathbf{b}\mathbf{b}^T}{2\psi}\;. 
  \end{align*}
  We can separately upper bound the eigenvalues of each term on the righthand side, obtaining the bound  
  \begin{align*}
    \lambda_n\left(\mathbf{B} + \frac{\mathbf{b}\mathbf{b}^T}{2\psi} \right)\leq \max_\ell\beta_\ell + \frac{1}{2\psi}\sum_\ell \beta_\ell^2 \leq n(n-1). 
  \end{align*} 
  In the final inequality we have used that $\beta_n \leq n-1$ and $\BETA \geq \e$. 
  Inverting this expression gives the first factor in the statement of the lemma. 

  Entrywise Taylor expansion, valid in $\mathcal{B}_\delta$, yields  
  \begin{align}
    \mathbf{A} + \mathbf{D} = (\mathbf{F} + \mathbf{G}_1)  + 2(\mathbf{F}^{\circ 2} + \mathbf{G}_2) + 3(\mathbf{F}^{\circ 3} + \mathbf{G}_3) + \cdots  \label{eq:matrix_series}
  \end{align}
  where $\mathbf{F} = \frac{\BETA\BETA^T}{2\psi}$, $\mathbf{F}^{\circ p}$ denotes Hadamard (entrywise) exponentiatian, and $\mathbf{G}_p$ is the diagonal matrix with entries
  \begin{align*}
    {[\mathbf{G}_p]}_{ii} = \sum_{j = 1}^n x_{ij}^p - 2x_{ii}^p\;.
  \end{align*}
  Each term in the series \eqref{eq:matrix_series} is weakly diagonally dominant, and therefore positive semi-definite by the Ger\u{s}gorin disc theorem \cite{horn2012matrix}. 
  It will therefore suffice to bound below the eigenvalues of the first term. 
  Since $\mathbf{F}$ is positive-semidefinite, we can obtain lower bounds in the eigenvalues by checking the eigenvalues of $\mathbf{G}_1$. 
  Each entry is 
  \begin{align*}
    [\mathbf{G}_1]_{ii} = \sum_{j = 1}^n\frac{\beta_i\beta_j}{2\psi} - 2\frac{\beta_i^2}{2\psi} = \beta_i - 2\frac{\beta_i^2}{2\psi} = \beta_i\left(1 - \frac{\beta_i}{\psi} \right)> 0\;. 
  \end{align*}
  We can bound this expression away from $0$ by noting that 
  \begin{align}
    \beta_i\left(1 - \frac{\beta_i}{\psi} \right) \geq \min \left\{1 - \psi^{-1}, \beta_n\left(1 - \frac{\beta_n}{\psi} \right)\right\}\;.\label{eq:min}
  \end{align}
  By hypothesis, $2\psi \geq n \geq 5$, and therefore $\psi \geq 5/2$. 
  The first argument of the minimum is thus no smaller than $\frac{3}{5}$.
  On the other hand, inserting the bound $\beta_n \leq \sqrt{2\psi - \delta} < \sqrt{2\psi}$, we can bound the second argument as 
  \begin{align*}
      \beta_n\left(1 - \frac{\beta_n}{\psi} \right) \geq 1 - \frac{2}{\sqrt{5}} > 0\;, 
  \end{align*} 
  Inserting this lower bound into \eqref{eq:min} yields the result. 
  
\section{Proof of \Cref{thm:properties}}\label{sec:SI_properties}

  It is convenient to define the function $c:[\beta_1,\beta_n]\rightarrow \R$ by  
  \begin{align*}
    c_{\BETA}(z) = \sum_{\ell} \frac{z\beta_\ell}{2\psi - \beta_\ell z} - \frac{z^2}{2\psi - z^2}\;.
  \end{align*}
  Note that $c_{\BETA}(\beta_i) = h_i(\BETA)$.
  Then, supposing that $\BETA$ is a solution to \eqref{eq:to_solve}, $c_{\BETA}(\beta_i) = d_i$. 
  Additionally, if $\BETA \in \mathcal{B}_\delta$, then $c$ is continuously differentiable on the interval $[\beta_1,\beta_n]$.
  We will therefore show that $c'_{\BETA}(z) > 0$ for all $z$ on this interval. 
  The derivative is
  \begin{align*}
    c'_{\BETA}(z) = 2\psi\left(\sum_{\ell} \frac{\beta_\ell}{(2\psi - \beta_\ell z)^2} - \frac{z}{(2\psi - z^2)^2}\right) \;.
  \end{align*}
  We will compute Taylor series of the terms inside the parentheses.
  Convergence of these series is guaranteed by the hypothesis that $\BETA \in \mathcal{B}_\delta$. 
  It is convenient to introduce the notation $\psi_p = \frac{1}{2} \sum_{\ell}\beta_\ell^p$. 
  Note that $\psi = \psi_1$. 

  First, 
  \begin{align*}
    \sum_{\ell}\frac{\beta_\ell}{(2\psi - \beta_\ell z)^2} = \frac{1}{4\psi^2}\sum_{\ell = 1}^n\beta_\ell\sum_{p = 1}^\infty p\left(\frac{\beta_\ell z}{2\psi}\right)^{p-1} = \frac{1}{4\psi^2}\sum_{p = 1}^\infty p\left(\frac{ z}{2\psi}\right)^{p-1}\left(2\psi_{p}\right)\;.
  \end{align*}
  Next, 
  \begin{align*}
    \frac{z}{(2\psi - z^2)^2} = \frac{1}{4\psi^2} \sum_{p = 1}^\infty p \left(\frac{z^2}{2\psi}\right)^{p-1} = \frac{1}{4\psi^2} \sum_{p = 1}^\infty p \left(\frac{z}{2\psi}\right)^{p-1} z^{p-1}\;.
  \end{align*}
  We thus write
  \begin{align*}
    c'_{\BETA}(z) = \frac{1}{2\psi}\sum_{p = 1}^{\infty} p\left(\frac{z}{2\psi}\right)^{p-1}\left(2\psi_p - z^{p-1} \right)\;.
  \end{align*}
  Since $z \leq \beta_n$ by hypothesis, $z^{p-1} \leq 2\psi_p$. 
  Each term in the ezpansion is therefore strictly positive and we conclude that $c'_{\BETA}(z) > 0$.
  This proves (a). 

  To prove (b), we truncate to second order, obtaining
  \begin{align}
    c'_{\BETA}(z) > \frac{1}{2\psi}\left(2\psi - 1 + \frac{z}{\psi}\left(2\psi_2 - z\right)\right)\;. \label{eq:g_prime_bound}
  \end{align}
  A small amount of algebra in combination with the hypotheses that $n \geq 5$ and $\e \leq \BETA \leq (n-1)\e$ shows that $\frac{z}{\psi}\left(2\psi_2 - z\right) \geq 1$ for $z \in [\beta_1,\beta_n]$. 
  Inserting this lower bound into \eqref{eq:g_prime_bound}, we obtain 
  \begin{align*}
    c'_{\BETA}(z) > 1\;.
  \end{align*}
  In particular, $c_{\BETA}^{-1}$ is a nonexpansive map. 
  Since $c_{\BETA}^{-1}:d_i\mapsto \beta_i$, claim (b) follows.   

  To prove (c), it suffices to show that $\beta_1 \leq d_1$ and apply (b).
  Since $\BETA \in \mathcal{B}_\delta$, $\beta_1\geq 1$. 
  We compute
  \begin{align*}
    d_1 = \beta_1 \sum_{\ell\neq 1}\frac{\beta_\ell}{2\psi - \beta_1\beta_\ell} 
    \geq \beta_1 \sum_{\ell\neq 1}\frac{\beta_\ell}{2\psi - \beta_1} 
    = \beta_1\;.
  \end{align*}
  The first inequality uses $\beta_\ell \geq \beta_1$ and the hypothesis $\beta_1 \geq 1$.

\end{document}


\section{Supplementary Information} \label{sec:supplementary}

	\subsection{Jacobian of $\mathbf{h}$}

		\todo{Requires touching up and notational revision}

		Let $h:\R^n\setminus \{\mathbf{0}\} \rightarrow \R^n$ be defined componentwise by the formula 
	\begin{align*}
		h_i(\BETA) = \sum_{j \neq i}\frac{\beta_i\beta_j}{2\psi - \beta_i \beta_j}\;.
	\end{align*}
	Let $\mathbf{J}$ be the Jacobian of $h$.
	We restrict attention to the feasible set 
	\begin{align*}
	 	\mathcal{B} = \{\BETA | \BETA > \mathbf{0}\;, \; \BETA \BETA^T > 2\psi \mathbf{E}\}\;.
	\end{align*} 
	Numerical evidence suggests the following conjecture: 
	\begin{conj}
		Let $\lambda_n(\BETA)$ be the smallest eigenvalue of $\mathbf{J}(\BETA)$. 
		Then, for all $\BETA \in \mathcal{B}$, $\lambda_n(\BETA) > 0$. 
		In particular, $\mathbf{J}$ is positive definite and nonsingular. 
		Furthermore, if $\BETA \geq 1$ entrywise, then $\lambda_n(\BETA) \geq 1$. 
	\end{conj}
	\begin{remark}
		The numerics also indicate that $\mathbf{J}_{ik}(\BETA)$ is not even weakly diagonally dominant, so we will need to rely on another strategy in order to prove this conjecture. 
	\end{remark} 

	The first thing to do when exploring this conjecture is of course to calculate $\mathbf{J}$.
	Let's try to do it in a somewhat more streamlined form. 
	Recall that $x_{ij} = \frac{\beta_i\beta_j}{2\psi}$ and $w_{ij} = \frac{x_{ij}}{1 - x_{ij}}$. 
	We compute
	\begin{align*}
		\frac{\partial x_{ij}}{\partial \beta_k} = 
		\begin{cases}
			-\frac{x_{ij}}{2\psi} &\quad k \neq i,j\\ 
			x_{ij}(\frac{1}{\beta_i} -\frac{1}{2\psi}) &\quad k = i\\ 
			x_{ij}(\frac{1}{\beta_j} -\frac{1}{2\psi}) &\quad k = j\;. 
		\end{cases}
	\end{align*}
	and
	\begin{align*}
		\frac{\partial w_{ij}}{\partial \beta_k} = \frac{1}{(1-x_{ij})^2} \frac{\partial x_{ij}}{\partial \beta_k}\;.
	\end{align*}
	We can therefore write

	\begin{align*}
		\mathbf{J}_{ik}(\BETA) = \frac{\partial h_i(\BETA)}{\partial \beta_k} = \sum_{j \neq i} \frac{\partial w_{ij}}{\partial \beta_k} = 
		\begin{cases}
		 	\left(\frac{1}{\beta_i} - \frac{1}{2\psi}\right)\sum_{j \neq i} \frac{x_{ij}}{(1-x_{ij})^2}
		 		&\quad k = i \\ 
		 		\frac{1}{\beta_k}\frac{x_{ik}}{(1-x_{ik})^2} - \frac{1}{2\psi}\sum_{j \neq i} \frac{x_{ij}}{(1-x_{ij})^2}&\quad k \neq i\;.
		\end{cases}
	\end{align*}
	It is convenient to define the matrix $\mathbf{A}\in \R^{n\times n}$ with entries $a_{ij} = \frac{x_{ij}}{(1-x_{ij})^2}$.
	Note that $\mathbf{A}$ is symmetric, because $\mathbf{X}$ is. 
	In this case, we can write 
	\begin{align*}
		\mathbf{J}_{ik}(\BETA) = \begin{cases}
		 	\left(\frac{1}{\beta_i} - \frac{1}{2\psi}\right)\sum_{j \neq i} a_{ij} &\quad k = i \\ 
		 		\frac{1}{\beta_k}a_{ik} - \frac{1}{2\psi}\sum_{j \neq i} a_{ij}&\quad k \neq i\;.
		\end{cases}
	\end{align*}
	Let $\mathbf{B} = \diag \BETA$. 
	The Jacobian can be written in matrix form as 
	\begin{align*}
		\mathbf{J}(\BETA) =  \mathbf{B}^{-1}\mathbf{A} - \frac{1}{2\psi}\mathbf{A}\mathbf{E} + \mathbf{B}^{-1}\mathbf{D}\;,
	\end{align*}
	where $\mathbf{D} = \diag \mathbf{A}\mathbf{e}$.

%% file: shared.tex
\usepackage{lipsum}
\usepackage{amsfonts}
\usepackage{amsmath}
\usepackage{amssymb}
\usepackage{MnSymbol}
\usepackage{graphicx}
\usepackage{epstopdf}
\usepackage{pc_math}
\usepackage{bbm}
\usepackage{todonotes}
\usepackage[ruled, vlined,algo2e,linesnumbered]{algorithm2e}
\SetKwComment{Comment}{$\triangleright$\ }{}

\usepackage{subfiles}
\usepackage{subcaption}
\usepackage{substr}

\ifpdf
  \DeclareGraphicsExtensions{.eps,.pdf,.png,.jpg}
\else
  \DeclareGraphicsExtensions{.eps}
\fi

\newsiamremark{remark}{Remark}
\newsiamremark{hypothesis}{Hypothesis}
\crefname{hypothesis}{Hypothesis}{Hypotheses}
\crefname{conj}{Conjecture}{Conjectures}
\crefname{algocf}{Algorithm}{Algorithms}
\crefname{thm}{Theorem}{Theorems}
\crefname{lm}{Lemma}{Lemmas}
\crefname{prop}{Proposition}{Propositions}
\newsiamthm{claim}{Claim}
\newsiamthm{dfn}{Definition}
\newsiamthm{thm}{Theorem}
\newsiamthm{lm}{Lemma}
\newsiamthm{conj}{Conjecture}
\newsiamthm{prop}{Proposition}

\headers{Moments of  Random Multigraphs}{Philip S. Chodrow}

\title{Moments of Uniform Random Multigraphs with Fixed Degree Sequences\thanks{Submitted \today \funding{PSC acknowledges support from the National Science Foundation under Graduate Research Fellowship grant 1122374.}}}

\author{Philip S. Chodrow\thanks{Operations Research Center and Laboratory for Information and Decision Systems, Massachusetts Institute of Technology 
  (\email{pchodrow@mit.edu}).}
}

\usepackage{amsopn}
\DeclareMathOperator{\diag}{diag}
\graphicspath{{fig/}{../fig/}}

\makeatletter
\newcommand*{\addFileDependency}[1]{
  \typeout{(#1)}
  \@addtofilelist{#1}
  \IfFileExists{#1}{}{\typeout{No file #1.}}
}
\makeatother

\newcommand*{\myexternaldocument}[1]{%
    \externaldocument{#1}%
    \addFileDependency{#1.tex}%
    \addFileDependency{#1.aux}%
}

%% file: tex/intro.tex
\section{Introduction}
	
	The language of graphs offers a standard formalism for representing systems of interrelated objects or agents. 
	Simple graphs model agents connected by a single, usually static, relation, such as acquaintanceship, proximity, or similarity. 
	In many data sets, however, agents are linked by multiple, discrete interactions.
	Two agents in a contact network may be in spatial proximity multiple times in the study period.
	Two agents in a communication network may exchange many emails over the course of a week. 
	In an academic collaboration network, the same two authors may be jointly involved in tens or even hundreds of papers. 
	In such cases, it is natural to draw a distinct edge between agents for  each interaction event. 
	Doing so results in a multigraph, in which any two nodes may be linked by an arbitrary, nonnegative, integer-valued number of edges. 

    A fundamental tool in network data science is null model comparison, which allows the analyst to evaluate whether a feature observed in a given network is surprising when compared to benchmark expectations. 
    We therefore often compare observed networks against random graph null models -- probability distributions over graphs. 
    An especially common class of null models is obtained by fixing the degree 
	sequence $\degree$ of the observed network, which encodes the number of interactions for each node. 
	The degree sequence is known to constrain many of a network's macroscopic properties  \cite{Newman2001}. 
   	The least informative (or entropy-maximizing) distribution so obtained is the uniform distribution on the space of graphs with the specified degree sequence. 
   	The same construction goes through for multigraphs. 
   	When studying interaction networks, the corresponding random graph is the uniform distribution $\eta_\degree$ on the set $\mathcal{G}_\degree$ of multigraphs with degree sequence $\degree$.  

    In many applications, a set of complete samples from $\eta_\degree$ is not required -- only some selected moments.
    An especially important set of moments is summarized by the expected adjacency matrix.   
	We therefore consider the following question: if $\W$ is the (random) adjacency matrix of multigraph $G \sim \uniform$, what is the value of the expected adjacency matrix $\Eomega \triangleq \E[\W]$?
	The entry $\omega_{ij}$ of $\Eomega$ gives the expected number of edges between nodes $i$ and $j$. 
	These moments have several important applications in network science. 
	Among these is community-detection via modularity-maximization \cite{Newman2006}, which in many formulations includes a term for the expected number of edges between nodes under a suitably specified null model. 
	Despite its simplicity and relevance for applications, this problem has received relatively little mathematical attention. 

	Before surveying existing approaches to the estimation of $\Eomega$, we fix some notation. 
	Let $\mathcal{G}_\degree$ refer to the set of multigraphs without self-loops with degree sequence $\degree \in \Z_{+}^n$.
	From a modeling perspective, the exclusion of self-loops reflects an assumption that agents do not meaningfully interact with themselves. 
	An element $G\in \mathcal{G}_\degree$ has a fixed number $n$ of nodes and $m = \frac{1}{2}\sum_{i}d_i$ of edges. 
	We use bold uppercase symbols  to denote matrices, bold lowercase symbols to denote vectors, and standard symbols to denote scalars.
	We do not notationally distinguish deterministic and random objects, instead relying on their associated definitions.  
	We use Greek letters to denote expectations of random objects.
	An estimator of a quantity, either deterministic or stochastic, is distinguished by a hat.
	For example, $\Eomega = \E[\W]$ is the expectation of $\W$. 
	An estimator of $\Eomega$, either deterministic or stochastic, may be written $\hat{\Eomega}$. 
    
    One approach to estimating $\Eomega$ is Monte Carlo sampling. 
	We sample $s$ independent and identically distributed samples $\W^{(1)},\ldots,\W^{(s)} \sim \uniform$, and construct the estimator
	\begin{align}
		\WMC(\degree) \triangleq \frac{1}{s}\sum_{\ell = 1}^s \W^{(\ell)}\;. \label{eq:MCMC_estimator}
	\end{align}
	The estimator $\WMC$ is a random function of $\degree$, parameterized by the sample size $s$. 
	The Strong Law of Large Numbers (SLLN) ensures that $\WMC\rightarrow \Eomega$ almost surely as the number of samples $s$ grows large.
	Stronger results are possible: since each entry $\wMC_{ij}$ is bounded, the variance of $\wMC_{ij}$ is finite and we can apply the Central Limit Theorem to provide quantitative bounds on the convergence rate. 
	This attractive picture is marred by a severe computational inconvenience: the size and complex combinatorial structure of $\mathcal{G}_d$ makes exact sampling intractable. 
	Markov Chain Monte Carlo (MCMC) methods \cite{Fosdick2018} are therefore required.
	MCMC introduces a new complication: for any finite number of iterations, the samples produced will always be statistically dependent, and may therefore over-represent some regions of $\mathcal{G}_\degree$ and under represent others.
	This dependence breaks the guarantees provided by the SLLN or Central Limit Theorem.  
	Control over the \emph{mixing time} of the sampler is in principle sufficient to ameliorate this issue; however, there are few known mixing time bounds on MCMC samplers of distributions on $\mathcal{G}_\degree$. 
	Available upper bounds on the mixing times \cite{greenhill2011polynomial, greenhill2014switch,erdHos2019mixing} are too large for guarantees in many practical computations, and there are heuristic reasons to believe that there are limits on our ability to improve these bounds.  
	
	An alternative estimator $\WNG$, extremely common in the network science literature, is defined entrywise by a simple formula: 
	\begin{align}
		\wNG_{ij}(\degree) = f_{ij}(\degree) \triangleq 
		\begin{cases}
			\frac{d_id_j}{2m} &\quad i \neq j \\ 
			0 &\quad i = j\;.
		\end{cases} \label{eq:NG}
	\end{align}
	The function $f_{ij}$ plays an important role throughout this article. 
	Unlike $\WMC$, $\WNG$ is a deterministic function of $\degree$ that is essentially free to compute.  
	The functional form of $f_{ij}(\degree)$ can be derived in multiple ways. 
	For example, it is the expected edge density between distinct nodes $i$ and $j$ in the model of Chung and Lu \cite{chung2002connected,Chung2002}, which preserves $\degree$ in expectation rather than deterministically.
	We will therefore refer to \eqref{eq:NG} as the ``CL estimate'' after Chung and Lu, though we emphasize that these authors did not use this expression as an estimator for any of the models we consider here, and indeed restricted their attention to graphs without parallel edges. 
	The estimator $\WNG$ was also derived heuristically by Newman and Girvan when they introduced modularity maximization as a method for community detection in networks \cite{Newman2003,Newman2006}. 
	In their derivation, we approximate the number of edges between $i$ and $j$ as follows. 
	Node $i$ has $d_i$ edges. 
	Each of these edges must connect to one of the $n-1$ other nodes. 
	A ``random edge'' is attached to node $j$ with probability roughly $\frac{d_j}{2m - d_i}$. 
	Assuming that $d_i \ll 2m$ yields $\wNG_{ij}$ as an approximation.
	It is important to note that this heuristic argument does not formalize any probability measure over a set of graphs.
	Thus, although $\WNG$ is sometimes described as the expectation of a ``random graph with fixed degree sequence,'' this is not exactly true for any common models except that of Chung and Lu, in which degrees are fixed only in expectation.
	In particular, $\WNG$ possesses no guarantees related to its performance as an estimator for the uniform model $\uniform$, the most literal mathematical operationalization of the phrase ``random graph with fixed degree sequence.''
	As we will see this performance can indeed be quite poor on data sets with high edge densities. 
    
	In this article, we construct an estimator of $\Eomega$ for dense multigraphs that is both scalable and accurate.
	By treating an MCMC sampler for $\uniform$ as a stochastic dynamical system whose state space is $\mathcal{G}_\degree$, we derive stationarity conditions describing the desired moments. 
	As we will show, there exists a vector $\BETA \in \R_+^n$ such that $\chi_{ij}$, the probability that $w_{ij} \geq 1$, is given by  
	\begin{align*}
		\chi_{ij}&\triangleq \uniform(w_{ij} \geq 1) \approx \frac{\beta_i\beta_j}{\sum_i \beta_i} = f_{ij}(\BETA)
	\end{align*}
	for all $i\neq j$. 
	The function $f_{ij}$ in this approximation is the same as that which appears in the definition of the CL estimator in \eqref{eq:NG}.  
	Furthermore, the entries of $\Eomega$ are given approximately by 
	\begin{align*}
		\omega_{ij} \approx \frac{\chi_{ij}}{1 - \chi_{ij}}\;.
	\end{align*}
	Taken to together, these two formulae provide a method for computing an estimate of $\Eomega$ given knowledge of the vector $\BETA$.
	We construct an estimator $\hat{\BETA}$ of this vector by solving the system of $n$ nonlinear equations  
	\begin{align*}
		\sum_{j} \frac{f_{ij}(\BETA)}{1 - f_{ij}(\BETA)} = d_i\;, \quad i = 1,\ldots,n\;.
	\end{align*}
	We show that the solution to this equation, provided it exists, is unique within the realm of interpretable sequences $\BETA$ subject to mild regularity conditions, and that this solution can be found efficiently by a simple, iterative algorithm. 
	From $\hat{\BETA}$ we construct an estimator $\WI$ of $\Eomega$.
	As we show, this estimator is both easier to compute than $\WMC$ and much more accurate than $\WNG$. 
	Furthermore, we can view the Chung-Lu estimator $\WNG$ as an approximation of $\WI$, obtained from the latter via a sequence of two linear approximations. 
	
	\subsection{Outline}
		In \Cref{sec:configuration_models}, we review two important null multigraph models -- the configuration model and the uniform model -- as well as a unified MCMC algorithm for sampling from each. 
		The analysis of this algorithm forms the heart of our derivation of the estimate $\WI$ in \Cref{sec:moments}. 
		This estimator depends on the unknown vector $\BETA$, which must be learned from $\degree$. 
		We offer a simple scheme for doing so in \Cref{sec:optimization}, including a qualified uniqueness guarantee on the resulting estimator $\hat{\BETA}$ of $\BETA$; a description of its structure; and a numerical scheme for computing it efficiently. 
		In \Cref{sec:results} we turn to experiments. 
		We first study the behavior of our methods on two synthetic data sets, including a bootstrap-style test of the conjecture underlying our error-bounds. 
		We then check the accuracy of $\WI$ on a subset of a high school contact network. 
		Whereas $\WNG$ is significantly biased on this data set,  $\WI$ is nearly unbiased and decreases the mean relative error of the estimate by an order of magnitude. 
		In our final experiment, we study the behavior of modularity maximization when the standard null expectation $\WNG$ is replaced by $\WI$. 
		We find that the behavior of a multiway spectral algorithm \cite{zhang2015multiway} depends strongly on both the choice of null expectation and the data set under study. 
		We close in \Cref{sec:discussion} with a discussion and suggestions for future work.

%% file: tex/random_graphs.tex
\section{Random Graphs with Fixed Degree Sequences} \label{sec:configuration_models}
	
	Our interest will focus on the uniform model $\uniform$, but it will be useful draw comparisons to the somewhat more commonly-used configuration model \cite{Bollobas1980}.
	\begin{dfn}[Configurations] \label{def:stub}
	    For a fixed node set $N$ and degree sequence $\degree \in \mathbb{Z}_+^n$, let
	    \begin{align*}
	      \Sigma_\degree = \biguplus_{i = 1}^n\left\{ i_1,\ldots,i_{d_i} \right\}\;,
	    \end{align*}
	    where $\uplus$ denotes multiset union.
        Thus, $\Sigma_\degree$ contains $d_i$ labeled copies of each node $i$. 
	    The copies $i_1,\ldots,i_{d_i}$ are called \emph{stubs} of node $i$.
	    A \emph{configuration} $C = (N,E)$ consists of the node set $N$ and an edge set $E$ which partitions $\Sigma_\degree$ into unordered pairs.
	    An edge in $E$ of the form $\{i_k,i_\ell\}$ is called a \emph{self-loop}. 
	    The process of forming $C$ from $\Sigma_\degree$ is often called \emph{stub-matching}.
	\end{dfn}
	Let $\mathcal{C}_\degree\subset \Sigma_\degree$ be the set of all configurations with degree sequence $\degree$ that do not include any self-loops. 
	There is a natural surjection $g: \mathcal{C}_\degree \rightarrow \mathcal{G}_\degree$. 
	The image of $C \in \mathcal{C}_\degree$ under $g$ is obtained by replacing all stubs with their corresponding nodes and consolidating the result as a multiset.
	The uniform distribution on $\mathcal{C}_\degree$ induces a distribution on $\mathcal{G}_\degree$ via $g$. 
	Denote by $g^{-1}:\mathcal{G}_\degree \rightarrow 2^{\mathcal{C}_\degree}$ the function that assigns to each element of $\mathcal{G}_\degree$ its preimage in $\mathcal{C}_\degree$ under $g$. 
	\begin{dfn}[Configuration Model]
	  	Let $\lambda_\degree$ be the uniform distribution on $\mathcal{C}_\degree$.
	    The \emph{configuration model} on $\mathcal{G}_\degree$ is the distribution $\configuration = \lambda_\degree\circ g^{-1}$.
	\end{dfn}
	The distinction between $\uniform$ and $\configuration$ -- and its implications for data analysis -- was recently highlighted by Fosdick et al. \cite{Fosdick2018}. 
	We have diverged from the terminology of the authors: our ``uniform model'' is their ``configuration model on non-loopy, vertex-labeled multigraphs'' and our ``configuration model'' is their ``configuration model on non-loopy, stub-labeled multigraphs.''

	The distinction between uniform and configuration models lies in how they weight graphs with parallel edges. 
	Let $C_1$ and $C_2$ be two configurations. 
	Suppose that $C_1$ contains the matchings $(i_1,j_1), (i_2, j_2)$ and $C_2$ contains the matchings $(i_1, j_2), (i_2,j_1)$, and that they otherwise agree on all other stubs. 
	Let $G = g(C_1) = g(C_2)$. 
	Under the uniform model, $G$ is considered to be a single state, weighted equally with all other states.
	Under the configuration model, on the other hand, the probability mass placed on $G$ is proportional to $\abs{g^{-1}(G)}$, reflecting  both $C_1$ and $C_2$ as distinct states. 
	In particular, the configuration model $\configuration$ will tend to place higher probabilistic weight on elements of $\mathcal{G}_\degree$ with large numbers of parallel edges than will the uniform model $\uniform$. 

	In the absence of parallel edges, the uniform and configuration models are closely related. 
    Let $A$ be the event that $G$ is simple, without self-loops or parallel edges. 
    Then, it is direct to show (e.g. \cite{Bollobas1980}) that, for all $G$, $\uniform(G|A) = \configuration(G|A)$. 
    The reason is that, when $G$ is simple, the sizes of the preimages $g^{-1}(G)$ depend only on the degree sequence $\degree$. 
    Since $\degree$ is fixed in $\mathcal{G}_\degree$, these preimages all have the same size.  
    Thus, when a \emph{simple} random graph is required, the uniform model $\uniform$ and configuration model $\configuration$ are in principle interchangeable, in the sense that we can sample from $\uniform(\cdot|A)$ by repeatedly sampling from $\configuration$ until a simple graph is produced. 
    Furthermore, when the degree sequence $\degree$ grows slowly relative to  $n$,  $\configuration(A)$ is bounded away from zero  by a function that depends on moments of $\degree$ when $n$ grows large \cite{Bollobas1980,molloy1995critical, Angel2016}. 
    This in turn provides an upper bound on the expected number of samples from $\configuration$ required to produce a single sample from $\eta_\degree(\cdot|A)$. 
    The computational importance of this relationship is that stub-matching for sampling from $\configuration$ is well-understood and often fast. 
 
    For dense graphs, $\configuration(A)$ may be extremely small, and the number of samples required to produce a simple graph may be prohibitive. 
    While it is possible to make post-hoc edits to the graph to remove self-loops and multiple edges \cite{Molloy1998,Sjostrand2019}, such methods can generate substantial and uncontrolled bias in finite graphs.  
    Second and more importantly for our context, there is no equivalence between the unconditional distributions $\uniform$ and $\configuration$ on spaces of multigraphs. 
    Stub-matching cannot therefore be used to sample from $\uniform$ when modeling considerations allow the presence of multiple edges. 
    
	\subsection{Markov Chain Monte Carlo}
	    An alternative approach to sampling uses Markov chains to explore structured sets of graphs. 
	    There exists a large constellation of related algorithms for this class of task, including the sampling of marginal-constrained binary matrices \cite{verhelst2008efficient,artzy2005generating};  degree-regular \cite{viger2005efficient,mckay1990uniform,jerrum1990fast} and degree-heterogeneous \cite{carstens2015proof,strona2014fast,blitzstein2011sequential,del2010efficient} simple graphs; and graphs with degree-correlation constraints \cite{amanatidis2015graphic}. 
	    Most of these algorithms operate by repeatedly swapping edges in such a way as to preserve the required graph structure. 
	    
	    A fairly general variant, formulated by Fosdick et al. \cite{Fosdick2018}, can sample from either the uniform model $\uniform$ or the configuration model $\configuration$ on $\mathcal{G}_\degree$. 
        We define an edge swap to be a random function of two edges that share no nodes.\footnote{Swaps involving edges that intersect are used when sampling from spaces that include self-loops \cite{Fosdick2018}.} 
        It interchanges a node on the first edge with a node on the second: 
        \begin{align*}
            \mathrm{EdgeSwap}((i,j), (k,\ell)) = 
            \begin{cases}
            (i,k), (j,\ell) &\quad \text{with probability $1/2$} \\ 
            (i,\ell), (j,k) &\quad \text{with probability $1/2$}\;.
            \end{cases}
        \end{align*}
        An edge swap does not change the total number of edges incident to nodes $i$, $j$, $k$, or $\ell$, and therefore preserves $\degree$. 
        Starting from a graph $G_0 \in \mathcal{G}_\degree$, repeated edge-swaps can therefore be used to obtain a random sequence of elements of $\mathcal{G}_\degree$. 
        Since each element of this sequence depends stochastically only on its predecessor, this sequence is a Markov chain. 
        We perform Markov Chain Monte Carlo as follows. 
        At each time step, we select two random edges $(i,j)$ and $(k,\ell)$, uniformly selected from the set of pairs of edges with four distinct node indices.  
        We then perform a pairwise edge-swap of these edges with \emph{acceptance probability}
        \begin{align}
            a((i,j), (k,\ell)) \triangleq 
            \begin{cases}
                1 &\quad \text{configuration model } \configuration \\ 
                (w_{ij}w_{k\ell})^{-1} &\quad \text{uniform model } \uniform\;.
            \end{cases} \label{eq:acceptance_probability}            
        \end{align}
        In the case that the edge-swap is not accepted, we record the current state again and resample. 
        Formally, 

        \begin{algorithm2e}[H]
           \DontPrintSemicolon
            \caption{MCMC Sampling for $\uniform$ and $\configuration$}\label{alg:MCMC}
            \KwIn{degree sequence $\degree$, initial graph $G_0 \in \mathcal{G}_{\degree}$, target distribution $\rho \in \{\uniform, \configuration\}$, sample interval $\delta t \in \mathbb{Z}_+$,  sample size $s \in \mathbb{Z}_+$.}
            \textbf{Initialization:} $t \gets 0$, $G \gets G_0$\;
            \For{$t =1,2,\ldots, s(\delta t)$}{
                sample $(i,j)$ and $(k,\ell)$ uniformly at random from $\binom{E_t}{2}$ \;
                \uIf{$\mathrm{Uniform}([0,1]) \leq a((i,j),(k,\ell))$}{
                 $G_{t} \gets \mathrm{EdgeSwap}((i,j),(k,\ell))$\;
                }
                \Else
                {
                $G_t \gets G_{t-1}$
                }
            }
            \KwOut{$\{G_t \text{ such that } t|\delta t\}$}
        \end{algorithm2e}
  	
  		For sufficiently large sample intervals $\delta t$, the output of \Cref{alg:MCMC} will be approximately i.i.d. according to the target distribution $\rho$, as guaranteed by the following result. 
        \begin{thm}[Fosdick et al. \cite{Fosdick2018}] \label{thm:MCMC}
            The Markov chain $\{G_t\}$ defined by \Cref{alg:MCMC} is ergodic and reversible with respect to the input distribution $\rho$. 
            As consequence, samples $\{G_t\}$ generated by \Cref{alg:MCMC} are asymptotically independent and identically distributed according to $\rho$ as $\delta t\rightarrow \infty$.
        \end{thm}
        
        These results provide a principled solution to the problem of asymptotically exact sampling from $\uniform$, and can therefore be used to construct an estimator $\WMC$ of $\Eomega$, given by \eqref{eq:MCMC_estimator}, with arbitrary levels of accuracy.
        It suffices to let the sample size $s$ and sample interval $\delta t$ grow large. 
        There are two performance-related issues when using \Cref{alg:MCMC} in practice, both of which are connected to the number of edges $m$.  
        First is the question of how large $\delta t$ should be to ensure that the samples are sufficiently close to independence. 
        Heuristically, $\delta t$ should scale with the mixing time of the chain, but very few bounds on mixing times for chains of this type appear to be available. 
        In several recent papers, Greenhill \cite{greenhill2014switch,greenhill2011polynomial} and collaborators \cite{erdHos2019mixing} have derived the only bounds known to this author for edge-swap Markov chains. 
        In the space of simple graphs, under certain regularity conditions on the degree sequence, they provide a mixing time bound with scaling $O({d_*}^{14}m^{10} \log m)$, where $d_* = \max_i d_i$. 
        The scaling of this upper bound very poor, especially with regard to $m$, and is therefore not reassuring for practical applications.   
        The second issue relates to the acceptance probabilities themselves. 
        In a dense multigraph the number $W_{ij}$ of edges between $i$ and $j$ will typically be large, resulting in low acceptance rates. 
        Indeed, supposing that a typical entry $W_{ij}$ scales approximately linearly with $m$, a typical acceptance probability would scale roughly as $m^{-2}$. 
        A standard coupon-collector argument shows that it takes roughly $O(m\log m)$ accepted transitions to ensure that each edge has been swapped at least once, which would appear a reasonable requirement for a well-mixed chain.  
        We therefore conjecture that the overall mixing time of \Cref{alg:MCMC} for the uniform model on dense multigraphs is no smaller than $O(m^3 \log m)$, though a more precise statement and proof would be welcome.  
        While much better than the best known proven results, such a scaling could likely be prohibitive for graphs of even modest size. 
        
        These considerations suggest that forming the MCMC estimate $\WMC$ may not be a computationally practical way to estimate $\Eomega$ when $m$ is large. 
        Despite these limitations, \Cref{alg:MCMC} lies at the heart of our main results in the next section. 

	    \IfSubStringInString{\detokenize{main}}{\jobname}{}{
			\bibliographystyle{plain}
			\bibliography{../references}
		}

%% file: tex/error_bounds.tex
\section{A Dynamical Approach to Model Moments} \label{sec:moments}
    
    We introduce some additional notation to facilitate calculations.  
	The transpose of vector $\mathbf{u}$ is denoted $\mathbf{u}^T$, and the inner product of $\mathbf{u}$ and $\mathbf{v}$ by $\mathbf{u}^T\mathbf{v}$. 
	We denote the $i$th row or column of matrix $\W$ by $\w_i$; all matrices we encounter will be symmetric and so no ambiguity will arise. 
	Let $\e$ be the vector of ones; the dimension of $\e$ will be clear in context. 
	Similarly, let $\e_i$ be the $i$th standard basis vector. 
	All sums over node indices $i,j,k,\ell$ have implicit limits from $1$ to $n$. 
	Finally, $a\wedge b$ and $a \vee b$ denote the pairwise minimum and maximum of scalars $a$ and $b$, respectively. 
    
    \Cref{alg:MCMC} describes a stochastic dynamical update on the space $\mathcal{G}_\degree$ of multigraphs, which we identify with the space of symmetric matrices with nonnegative integer entries and zero diagonals. 
    Let $\DELTA(t) = \W(t+1) - \W(t)$ be the (random) increment in $\W$ in timestep $t+1$. 
    We implicitly regard $\W$ and $\DELTA$ as functions of $t$, suppressing the argument for notational sanity when there is no possibility of confusion. 
    We can separate $\DELTA = \DELTA^+ - \DELTA^-$, where $\Delta^+_{ij} = (\Delta_{ij} \vee 0)$ and $\Delta^-_{ij} = (-\Delta_{ij} \vee 0)$. 
    The first term $\Delta_{ij}^+$ describes the (random) number of edges flowing into the pair $(i,j)$ and the second term $\Delta_{ij}^-$ the random number of edges flowing out.  
    Conservation of edges implies that $\sum_{ij} \Delta_{ij}^+ = \sum_{ij}\Delta_{ij}^-$. 
    Since a pair of nodes can only gain or lose one edge at a time under the dynamics, the entries $\Delta_{ij}^+$ and $\Delta_{ij}^-$ are Bernoulli random variables. 
    These Bernoulli variables are not independent, since at most two entries of each matrix are nonzero in a given timestep. 
    Let $\edelta^+ = \E[\DELTA^+]$ and $\edelta^- = \E[\DELTA^-]$. 

    Two things must hold at stationarity of \Cref{alg:MCMC}. 
    First, all moments of $\W$ must be constant in time. 
    Second, since the stationary distribution of \Cref{alg:MCMC} is the target distribution $\rho$ by construction, these moments of $\W$ are the desired moments of $\rho$. 
    We can therefore approximately compute moments of $\rho$ by approximately solving conveniently chosen stationarity conditions.  
	A useful set is given by 
	\begin{align}
		\E[w_{ij}(t+1)^p - w_{ij}(t)^p] = \E\left[\left(w_{ij}(t) + \Delta_{ij}(t)\right)^p - w_{ij}(t)^p\right]  = 0\;, \label{eq:equilibrium}
	\end{align}
	for positive integers $p$.
	These equations express directly the time-invariance of the moments $\E[w_{ij}(t)^p]$ at stationarity. 

\subsection{Illustration: The Configuration Model} \label{subsec:illustration}
    
    We will derive a version of the Chung-Lu estimator $\WNG$ for the configuration model by studying \eqref{eq:equilibrium} when $p = 1$. 
	  
	\begin{thm} \label{thm:stub}
	    Under the configuration model $\configuration$, for all $i\neq j$, 
	    \begin{align}
	        \omega_{ij} = \frac{d_id_j - \E[\w_i^T\w_j] - \E[w_{ij}^2]}{2m-d_i-d_j} \label{eq:configuration}
	    \end{align}
	\end{thm}
	\begin{proof}
	    We first derive expressions for $\DELTA^-$ and $\DELTA^+$ by stepping through the stages of \Cref{alg:MCMC}. 
	    For the former, note that $\Delta^-_{ij} = 1$ only if edge $(i,j)$ is sampled in the first stage of the iteration. 
	    The probability that edges $(i,j)$ and $(k,\ell)$ are sampled, assuming that all four indices are distinct, is $z(\W)^{-1}W_{ij}W_{k\ell}$, where 
	    \begin{align*}
	    	z(\W) = \sum_{\substack{i,j \\ k,\ell \notin \{i,j\}}} W_{ij}W_{k\ell}
	    \end{align*}
	    gives the total number of ways to pick two edges with four distinct indices. 
    	Under the configuration model, $a((i,j), (k,\ell)) = 1$. 
    	Summing across $k$ and $\ell$ and taking expectations, we obtain 
    	\begin{align*}
    		\delta^-_{ij} &= \frac{1}{z(\W)}\E\left[\sum_{\substack{k,\ell \notin \{i,j\}}} w_{ij}w_{k\ell}\right] \\ 
    		&= \frac{1}{z(\W)}\E\left[w_{ij} \left(\sum_{k,\ell} w_{k\ell} - \sum_{k}\left(w_{ki}+ w_{kj}\right) - \sum_{\ell}\left(w_{i\ell} + w_{j\ell}\right) + 3w_{ij} \right)\right]\;.
    	\end{align*}
    	Recalling constraints such as $\sum_{k,\ell}W_{k\ell} = 2m$ and $\sum_{k}W_{ki} = d_i$, this expression simplifies to
    	\begin{align*}
    		\delta^-_{ij} &= \frac{1}{z(\W)}\left(2\omega_{ij}(m - d_i - d_j) +3\E[w_{ij}^2] \right)
    	\end{align*}
    	We can derive a similar expression for $\delta_{ij}^+$. 
    	Fix two additional indices $k$ and $\ell$, such that all four indices $i,j,k,\ell$ are distinct. 
    	A new edge $(i,j)$ can be generated from selecting for swap either of the pairs $\{(i,k),(\ell,j)\}$ or $\{(i,\ell),(k,j)\}$. 
    	These events occur with probabilities $z(\W)^{-1}w_{ik}w_{\ell j}$ and $z(\W)^{-1}w_{i\ell}w_{k j}$, respectively. 
    	Having selected edges $\{(i,k), (\ell,j)\}$, edges $\{(i,j),(k,\ell)\}$ are formed by the swap with probability $\frac{1}{2}$; otherwise $\{(i,\ell), (k,j)\}$ are formed. 
    	Summing across $k$ and $\ell$ and computing expectations, we have 
    	\begin{align*}
    	    \delta^+_{ij} &= \frac{1}{2z(\W)}\E\left[\Sum_{\substack{k,\ell \notin \{i,j\} \\ k \neq \ell}} w_{ik}w_{\ell j} + \Sum_{\substack{k,\ell \notin \{i,j\} \\ k \neq \ell}} w_{i\ell}w_{k j}\right] \\
    	    &= \frac{1}{z(\W)}\E\left[\Sum_{\substack{k,\ell \notin \{i,j\} \\ k \neq \ell}} w_{ik}w_{\ell j}\right] \\
    	    &= \frac{1}{z(\W)} \E\left[\left(\sum_{k}w_{ik}\right)\left(\sum_\ell w_{\ell j}\right) - w_{ij}\sum_{k}\left(w_{ik} + w_{jk}\right) - \sum_k w_{ik}w_{kj} + w_{ij}^2\right] \\ 
    	    &= \frac{1}{z(\W)} \left(d_id_j\ - \omega_{ij}(d_i + d_j) - \E[\w_i^T\w_j] + \E[w_{ij}^2]\right)\;.
    	    \label{eq:delta_plus_stub}
    	\end{align*}
    	Choosing $p = 1$ in \eqref{eq:equilibrium}, we must have $\delta_{ij}^+ = \delta_{ij}^-$ at stationarity. 
    	Inserting our derived expressions and solving for $\omega_{ij}$ yields the result. 
	\end{proof}
	
    \Cref{thm:stub} does not give an explicit operational solution for $\omega_{ij}$, since the righthand side contains higher moments of $\W$. 
	Progress can be made in the ``large, sparse regime,'' in which  we assume  that $n$ is large and the entries of $\W$ and $\degree$ small relative to $m$. 
	Recalling that $\wNG_{ij} = \frac{d_id_j}{2m}$, we can rewrite  \eqref{eq:configuration} as  
	\begin{align*}
		\omega_{ij} &= \left(1 - \frac{d_i + d_j}{2m}\right)^{-1}\left(1 - \frac{\E[\w_i^T\w_j] - \E[w_{ij}^2]}{d_id_j}\right)\wNG_{ij}. 
	\end{align*}
	In the large, sparse heuristic, each entry of $\degree$ is small in comparison to $m$, and the first error factor is near unity. 
	Similarly, the expression $\E[\w_i^T\w_j] - \E[w_{ij}^2]$ implicitly contains up to $n-1$ nonzero products of entries of $\W$. 
	On the other hand, the denominator contains $(n-1)^2$ such terms, and we therefore expect the second error factor to also lie near unity. 
	We therefore expect that $\omega_{ij} \rightarrow \wNG_{ij}$ ``in the large, sparse regime.''. 
	This statement can be made precise by specifying the asymptotic behavior of $\W$ with respect to $n$, which is beyond our present scope.  
	Through analysis of \Cref{alg:MCMC}, we have derived both $\WNG$ and explicit error terms that are often elided in the network science literature. 

\subsection{Moments of the Uniform Model}

	The analysis of the uniform model is somewhat more subtle. 
	As seen in the configuration model above, many of the sums that appear in the calculations of $\edelta^+$ and $\edelta^-$ reduced to fixed constants due to the degree constraints. 
	Unfortunately, there is no analogous simplification in the uniform model $\uniform$. 
	Because of this, we require some additional technology in order to make progress. 

	Define the binary matrix $\X \in \{0,1\}^{n\times n}$ entrywise by $x_{ij} = \mathbbm{1}(w_{ij} = 0)$.
	For convenience, we adopt the convention $0/0 = 0$ under which the identity $w_{ij}/w_{ij} = x_{ij}$ holds even when $w_{ij} = 0$. 
	We can interpret $\X$ as the adjacency matrix of the simple graph obtained by collapsing all sets of parallel edges in a multigraph into single edges.  
	Let $\b = \X\e$, the vector of row sums of $\X$. 
	The vector $\b$ is interpretable as the collapsed degree sequence, whose $i$th entry gives the number of distinct neighbors of node $i$.
	Let $y = \frac{1}{2}\e^T\b$ give the total number of collapsed edges. 
	The expectations of $\X$, $\b$, and $y$ play important roles in our analysis. 
	We denote them 
	\begin{align*}
		\CHI = \E[\X]\;, \quad \BETA = \E[\b]\;, \text{ and} \quad \psi = \E[y]\;.
	\end{align*}
	The objects $\CHI$,  $\BETA$, and $\psi$ are all implicitly deterministic functions of $\degree$.
	Throughout this section, we let $I = (i_1,\ldots,i_p)$ be a set of $p$ not-necessarily-distinct  indices, and let $K = ((k_1,\ell_1),\ldots,(k_q,\ell_q))$ be a set of $q$ not-necessarily-distinct dyadic indices. 
	If $\mathbf{v}$ is a vector, we let $\mathbf{v}_{I}$ denote the vector with entries $(v_{i_1},\ldots,v_{i_p})$. 
	Similarly, if $\mathbf{A}$ is a matrix, we let $\mathbf{A}_{K}$ denote the vector with entries $(a_{k_1,\ell_1},\ldots,a_{k_q,\ell_q})$.
	
	We first require control over the behavior of $\BETA$ with respect to $\degree$.
	\begin{dfn}[Regularity Constant for $\BETA$]
		Let $u(\degree)$ be the smallest real number such that, for any degree sequence $\degree'$ such that $\degree'\geq \degree$ entrywise, and for all distinct indices $i$ and $j$, 
		\begin{align*}
			\norm{\BETA(\degree' + \e_i + \e_j) - \BETA(\degree')}_{\infty} \leq u(\degree)\;.
		\end{align*}
	\end{dfn}
	Intuitively, $u(\degree)$ provides a bound on the sensitivity of $\BETA$ to increments in entries of the degree sequence. 
	Indeed, $u(\degree)/\sqrt{2}$ is by definition the Lipschitz constant (with respect to the $\ell^2$ and $\ell^\infty$ norms)  for the restriction of $\BETA$ to the set $\{\degree':\degree'\geq \degree\}$. 
	Since $0\leq \beta_\ell(\degree) \leq n-1$, we have trivially that $u(\degree) \leq n-1$. 
	To produce a lower bound, we can also produce degree sequences $\degree$ such that $u(\degree) \geq 1$. 
	To see this, note that, if $d_i = d_j = 0$, then $\beta_i(\degree) = \beta_j(\degree) = 0$. 
	On the other hand, $\beta_i(\degree + \e_i + \e_j) = \beta_j(\degree + \e_i + \e_j) = 1$.

	We also define $v$ as the smallest real number such that, for all $i$ and $j$,  
	\begin{align*}
		\abs{\psi(\degree + \e_i + \e_j) - \psi(\degree)}\leq  v(\degree)\;.
	\end{align*}
	Similarly to the above, we have the trivial bounds $1 \leq v(\degree) \leq n(n-1)$, since  
	$n(n-1)$ is the largest possible number of nonzero entries of $\X$. 
	\begin{conj} \label{conj:u}
		For all $\degree$, we have  $u(\degree) \leq 1$ and $ v(\degree) \leq 1$.
	\end{conj}
	The intuition behind this conjecture is as follows. 
	If $G \in \mathcal{G}_\degree$, the sequence $\degree + \e_i + \e_j$ can be instantiated by a graph $G' \in \mathcal{G}_{\degree + \e_i + \e_j}$ in which a single edge has been added between nodes $i$ and $j$. 
	Trivially, this operation does not decrease the degrees of any nodes, does not increase the degrees of any nodes by more than one, and does not increase the total number of edges by more than one. 
	\Cref{conj:u} states that the same is true of the expected collapsed degrees $\BETA$ and expected number of collapsed edges $\psi$. 
	The bounds we present below do not formally depend on the truth of \Cref{conj:u}, but they are not guaranteed to be meaningful unless $u$ and $v$ are indeed small in comparison to $n$.
	Unfortunately, the complex combinatorial structure of $\mathcal{G}_\degree$ renders a proof of our conjecture obscure, and we leave such a proof to future work.  
	In \Cref{subsec:results}, we will show anecdotal numerical experiments consistent with this conjecture. 
	For now, we define 
	\begin{align*}
		u_*(\degree) = \max\{u(\degree),1\} \quad \text{and} \quad 	v_*(\degree) = \max\{v(\degree),1\}\;.
	\end{align*} 
	\Cref{conj:u} then states that $u_*(\degree) = v_*(\degree) = 1$ for all $\degree$. 

	\begin{thm} \label{thm:beta}
		Suppose that $\eta_{\degree}(\X_K = \mathbf{e}) > 0$. 
		Then, for any $i \in [n]$,
		\begin{align}
			 \abs{\E[b_{i}|\X_{K} = \e] - \beta_i(\degree)} \leq 2qu_*(\degree)\label{eq:beta}\;.
		\end{align}
		Additionally, 
		\begin{align}
			 \abs{\E[y|\X_{K} = \e] - \psi(\degree)} \leq 2qv_*(\degree)\;.\label{eq:psi}
		\end{align}
	\end{thm} 
	\begin{proof}
		We will prove \eqref{eq:beta}; the proof of \eqref{eq:psi} is parallel. 
		Fix $k,\ell \in [n]$. 
		The distribution $\eta_{\degree + \e_k + \e_\ell}$ is supported on graphs with $n$ nodes and $m + 1$ edges. 
		Let us condition $\eta_{\degree + \e_k + \e_\ell}$ on the event $w_{k\ell} \geq 1$. 
		Then, there exists at least one edge $(k,\ell)$. 
		Since $\eta_{\degree + \e_k + \ell}$ is itself uniform, the conditioned distribution, in which the remaining $m$ edges, is also uniform. 
		Indeed, since we have already assigned an edge incident to nodes $k$ and $\ell$, the conditional distribution is uniform over configurations of the remaining $m$ edges in which the degrees sum to $\degree$.
		This is exactly $\eta_\degree$, and we therefore obtain the identity 
		\begin{align}
			\eta_{\degree}(G) = \eta_{\degree + \e_k + \e_\ell}(G\uplus\{(k,\ell)\} |w_{k\ell} \geq 1)\;,\label{eq:conditioning}
		\end{align}
		where $\uplus$ denotes multiset union. 
		This identity relates the operations of conditioning and degree sequence modification. 
		Now, let $\mathbf{v}(K) = \sum_{s = 1}^q(\e_{k_s} + \e_{\ell_s}) $.
		By iterating \eqref{eq:conditioning}, we obtain 
		\begin{align*}
			\eta_\degree(G) = \eta_{\degree + \mathbf{v}(K)}\left(G \uplus  \biguplus_{s = 1}^q\{k_s,\ell_s\}\Big|\W_K \geq \e\right)\;. \label{eq:multi_conditioning}
		\end{align*}
		Note that the conditioning event can be equivalently written $\X_K = \e$. 

		For the remainder of this proof, let the symbol $\E_{\mathbf{z}}$ denote expectations with respect to $\eta_{\mathbf{z}}$. 
		We use \eqref{eq:conditioning} to estimate $\E_{\degree + \mathbf{v}(K)}[b_i|\X_K = \mathbf{e}]$ via a two-step experiment. 
		We first sample $G \sim \eta_{\degree}$ and compute $b_i$.
		We then add the $q$ edges $\{(k_s,\ell_s)\}$ sequentially. 
		Doing so does not decrease $b_i$, and can increase $b_i$ by no more than $q \leq qu_*$. 
		Taking expectations, we obtain the bound 
		\begin{align*}
			\beta_i(\degree) \leq \E_{\degree + \mathbf{v}(K)}[b_i|\mathbf{X}_K = \mathbf{e}] \leq \beta_i(\degree) + qu_*\;.
		\end{align*}
		Applying \eqref{conj:u} inductively, we also have 
		\begin{align*}
			\beta_i(\degree) - qu_*\leq \beta_i(\degree + \mathbf{v}(K)) \leq \beta_i(\degree) + qu_*\;.
		\end{align*}
		We infer that
		\begin{align*}
			\abs{\E_{\degree + \mathbf{v}(K)}[b_i|\X_K = \mathbf{e}] - \beta_i(\degree + \mathbf{v}(K))} \leq 2qu_*\;.
		\end{align*}
		Since $\degree$ and $K$ were arbitrary, we can absorb $\mathbf{v}(K)$ into $\degree$, obtaining the required statement 
		\begin{align*}
			\abs{\E_{\degree }[b_i|\X_K = \mathbf{e}] - \beta_i(\degree)} \leq  2qu_*\;,
		\end{align*}
		The only subtlety in this case is that the expectation must exist. 
		For this it is sufficient that $\eta_{\degree}(\X_K = \e)> 0$, as assumed by hypothesis. 
	\end{proof}

	\Cref{thm:beta} is our primary tool for proving second-moment bounds on the entries of $\mathbf{b}$.  
	From this point forward, we will assume that $\degree$ is fixed. 
	The symbols $\eta$, $\E$ will refer to the uniform model with degree sequence $\degree$ and expectations with respect to that model, respectively. 
	
	\begin{lm}
		Let $i\neq j \neq k$. 
		The following bounds hold. 
		\begin{align}
			\abs{\E[b_ix_{jk}] - \beta_i \chi_{jk}} &\leq 
				2u_*\chi_{jk}  
			\label{eq:basic} \\ 
			\abs{\E[yx_{jk}] - \psi \chi_{jk}} &\leq 
				2v_*\chi_{jk} 
			\label{eq:basic_2} \\ 
			\abs{\E[b_ib_j] - \beta_i\beta_j} &\leq 2u_*(\beta_i\wedge\beta_j) \label{eq:ij} \\ 	
			\E[b_i^2b_j^2]  &\leq (\beta_i + 6u_*)^2(\beta_j + 6u_*)^2 \label{eq:iijj} \\ 
			\var{y}   &\leq 2v_*\psi\;. \label{eq:var_y}
		\end{align}
	\end{lm}
	\begin{proof}
		To prove \eqref{eq:basic}, write 
		\begin{align*}
			\E[b_ix_{jk}] = \eta(x_{jk} = 1)\E[b_i|x_{jk} = 1] = \chi_{jk}\E[b_i|x_{jk} = 1]
		\end{align*}
		and apply \Cref{thm:beta}. 
		To prove \eqref{eq:basic_2}, we similarly write 
		\begin{align*}
			\E[yx_{jk}] = \eta(x_{jk} = 1)\E[y|x_{jk} = 1] = \chi_{jk}\E[y|x_{jk} = 1]
		\end{align*}
		and apply \Cref{thm:beta}. 
		To prove \eqref{eq:ij}, write
		\begin{align*}
			\E[b_ib_j] &= \sum_{\ell}\E[b_ix_{j\ell}]  
		\end{align*}
		Now applying \eqref{eq:basic}, we obtain
		\begin{align*}
			\abs{\E[b_ib_j] - \beta_i\beta_j} &\leq 2u_*\sum_{\ell} \chi_{j\ell} = 2u_*\beta_j\;.
		\end{align*}
		Since we could have expanded $b_j$ instead of $b_i$, we can choose the smaller of these, and the result follows. 
		The proof of \eqref{eq:iijj} is similar. 
		We expand the sums and apply \Cref{thm:beta}. 
		The first step is
		\begin{align*}
			\E[b_i^2b_j^2] &=\sum_{k,\ell,h} \E[x_{ik}x_{i\ell}x_{jh}]\E[b_j|x_{ik}x_{i\ell}x_{jh} = 1] \\ 
			&\leq \sum_{k,\ell,h} \E[x_{ik}x_{i\ell}x_{jh}](\beta_j + 6u_*) \\ 
			&= (\beta_j + 6u_*) \E[b_i^2b_j]\;.
		\end{align*}
		Repeating this procedure three more times proves the result. 
		Finally, to prove \eqref{eq:var_y}, write 
			\begin{align*}
				\var{y} &= \E[y^2] - \psi^2 = \frac{1}{2}\sum_{ij}\chi_{ij}\E[y|x_{ij} = 1] -\psi^2 \leq \frac{1}{2}\sum_{ij}\chi_{ij} (\psi + 2v_*)  - \psi^2 = 2v_*\psi\;.
			\end{align*}
			We have used \eqref{eq:basic_2} in the inequality.
		\end{proof}
 	
		\begin{lm}\label{lm:delta}
			We have 
			\begin{align*}
					\abs{\delta_{ij}^- - \frac{2\psi \chi_{ij}}{z(\W)}} \leq \frac{\epsilon_{ij}^-}{z(\W)} \quad \text{and} \quad 
					\abs{\delta_{ij}^+ - \frac{\beta_i\beta_j}{z(\W)}} \leq \frac{\epsilon_{ij}^+}{z(\W)}\;,
			\end{align*}		
			where 
			\begin{align*}
				\epsilon_{ij}^- &\triangleq \chi_{ij}(\beta_i + \beta_j + 3 + 4v_* + 2u_*) \\ 
				\epsilon_{ij}^+&\triangleq \chi_{ij}(\beta_i + \beta_j + 4u_* - 1) + (2u_* + 1)(\beta_i \wedge \beta_j)\;.
			\end{align*}
		\end{lm}
		\begin{proof}
			We require expressions for $\DELTA^-$ and $\DELTA^+$. 
		    These are as in the calculation for the configuration model in \Cref{subsec:illustration}, except that there now appears an acceptance probability $a((i,j),(k,\ell)) = \frac{1}{w_{ij}, w_{k\ell}}$ that modifies the swap probabilities. 
		    The acceptance probability has the effect of replacing instances of $w_{ij}$ with $x_{ij}$. 
			Performing the algebra and simplifying, we find that 
			\begin{align}
				\delta_{ij}^- 
			    &= \frac{1}{z(\W)} \E\left[2x_{ij}(y - b_i - b_j) + 3x_{ij}\right] \label{eq:delta_minus} \\
			    \delta^+_{ij} 
			    &= \frac{1}{z(\W)}\E\left[b_ib_j - x_{ij}(b_i + b_j) - \x_i^T \x_j + x_{ij}\right] \;. \label{eq:delta_plus}
			\end{align}
			These are indeed the same expressions as in the configuration model in \Cref{subsec:illustration}, with $\W$ replaced by $\X$. 
			We have used the identity $x_{ij}^2 = x_{ij}$. 
		    Computing the expectation of the first line  yields
		    \begin{align*}
		  		\delta_{ij}^- = \frac{1}{z(\W)}\left(2\E[yx_{ij}] - 2\E[b_ix_{ij}] - 2\E[b_jx_{ij}] + 3\chi_{ij}\right)\;.
		    \end{align*}
		    Applying \eqref{eq:basic} and \eqref{eq:basic_2}, we obtain the bound 
		    \begin{align*}
		    	\abs{\delta_{ij}^- - \frac{1}{z(\W)}\left(\chi_{ij}(2(\psi - \beta_i -\beta_j) + 3\right)} \leq \frac{4}{z(\W)}\chi_{ij}(v_* + 2u_*) \;.
		    \end{align*}
		    We similarly compute 
		    \begin{align*}
		    	\delta_{ij}^+ = \frac{1}{z(\W)}\left(\E[b_ib_j] - \E[b_ix_{ij}] - \E[b_jx_{ij}] - \E[\x_i^T\x_j] + \chi_{ij}\right)\;.		
		    \end{align*}
		    We note that, since $\X$ is binary, $0 \leq \x_i^T\x_j\leq b_i \wedge b_j$, and therefore $0 \leq \E[\x_i^T\x_j] \leq \beta_i \wedge \beta_j$.
		    Applying this observation in concert with \eqref{eq:ij} and \eqref{eq:basic}, we find 
		    \begin{align*}
		    	\abs{\delta_{ij}^+ - \frac{1}{z(\W)}\left(\beta_i\beta_j - \chi_{ij}\beta_i -\chi_{ij}\beta_j + \chi_{ij}\right)} \leq \frac{1}{z(\W)}\left((2u_* + 1)(\beta_i\wedge \beta_j) + 4u_* \chi_{ij} \right)\;.
		    \end{align*}
		    Moving the unwanted terms to the righthand side in both bounds proves the lemma. 
		\end{proof}
		\begin{thm}[Expectations of $\X$] \label{thm:X}
			We have 
			\begin{align*}
				\abs{\chi_{ij} - \frac{\beta_i\beta_j}{2\psi}} &\leq \epsilon_{ij}(\BETA) \triangleq \frac{\epsilon_{ij}^+(\BETA) + \epsilon_{ij}^-(\BETA)}{2\psi}\;.
			\end{align*}
			Furthermore, 
			\begin{align*}
				\epsilon_{ij}(\BETA) = \frac{2\chi_{ij}(\beta_i + \beta_j + 3u_* + 2v_* + 2) + (2u_*+1)(\beta_i\wedge\beta_j)}{2\psi}\;.
			\end{align*}
		\end{thm}
		\begin{proof}
			Setting $p = 1$ in \eqref{eq:equilibrium} again yields $\delta_{ij}^+ - \delta_{ij}^- = 0$. 
			Applying \Cref{lm:delta} and the triangle inequality, we obtain 
			\begin{align*}
				\abs{\frac{\beta_i\beta_j}{z(\W)} - \frac{2\psi\chi_{ij}}{z(\W)}} \leq \frac{\epsilon_{ij}^- + \epsilon_{ij}^+}{z(\W)} 
			\end{align*}
			Multiplying through by $\frac{z(\W)}{2\psi}$ proves the first claim.
			The expression for $\epsilon_{ij}$  is obtained inserting the expressions for $\epsilon_{ij}^-$ and  $\epsilon_{ij}^+$ from \Cref{lm:delta} and simplifying. 
		\end{proof}
		\Cref{thm:X} provides an asymptotic error bound of the form 
		\begin{align}
			\chi_{ij} = \frac{\beta_i\beta_j}{2\psi}\left(1 + O\left(\frac{\chi_{ij}v_*}{\beta_i\beta_j} + \chi_{ij}\frac{\beta_i + \beta_j}{\beta_i\beta_j} +\frac{ u_*}{\beta_i\vee \beta_j}\right)\right) \label{eq:x_bounds}
		\end{align}
		as $\beta_i$ and $\beta_j$ grow large. 
		This bound is admittedly relatively loose, even assuming that $u_*$ and $v_*$ are indeed small.
		In light of the numerical results presented below, we conjecture that much better bounds may be possible. 
		This appears to be a promising direction for future work. 

		We can recognize the leading term in \eqref{eq:x_bounds}:  
		\begin{align*}
			f_{ij}(\BETA) = \frac{\beta_i\beta_j}{2\psi}\;,
		\end{align*}
		the same functional form $f_{ij}$ as in the CL estimator defined in \eqref{eq:NG}. 
		Speaking somewhat figuratively, we can interpret \Cref{thm:X} as indicating that $\X$, the matrix of the projected simple graph, approximately agrees in expectation with the Chung-Lu model (on off-diagonal entries) with parameter vector $\BETA$. 
		However, tt would be incorrect to state that $\X$ is distributed according to any model that deterministically preserves a collapsed degree sequence. 
		First, $\BETA$ does not in general possess integer entries. 
		Second the collapsed degrees $b_i$ are still stochastic, preserved only approximately in expectation. 

	\subsection{First Moments of $\W$}
		
		In the case of the configuration model, approximately solving the $p = 1$ stationarity condition yielded an approximation for $\Eomega$ in terms of the known vector $\degree$. 
		However, in the uniform model we derived an approximation only for $\CHI$ in terms of the unknown vector $\BETA$. 
		Computing another equilibrium condition will allow us to both estimate $\Eomega$ from $\CHI$ and estimate $\BETA$ from $\degree$. 
		Take $p = 2$ in \eqref{eq:equilibrium}, obtaining
		\begin{align}
			2\E[w_{ij}\Delta_{ij}] + \E[\Delta_{ij}^2] = 0\;. \label{eq:p=2}
		\end{align}
		Study of this condition yields the following result. 
		\begin{thm} \label{thm:W}
			Assume that $f_{ij}(\BETA) < 1$. 
			Then, 
			\begin{align*}
				\abs{\omega_{ij} - \frac{f_{ij}(\BETA)}{1 - f_{ij}(\BETA)}} \leq \frac{1}{1-f_{ij}(\BETA)} \left(\frac{2\epsilon'_{ij}(\BETA) + \epsilon_{ij}(\BETA)}{2\psi} + \frac{\epsilon_{ij}(\BETA)}{2}\right)\;,
			\end{align*}
			where $\epsilon_{ij}(\BETA)$ is as in \Cref{thm:X} and 
			\begin{align*}
				\epsilon'_{ij}(\BETA) &\triangleq \frac{2u_*}{\beta_i\vee\beta_j} + \frac{\sigma_{ij}}{\omega_{ij}}\frac{\sqrt{(\beta_i + 6u_*)^2(\beta_j+6u_*)^2 - (\beta_i\beta_j - 2u_*(\beta_i\wedge\beta_j))^2}}{\beta_i\beta_j} \\ 
				&\quad  + \sigma_{ij}\sqrt{2v_*\psi_i} + \omega_{ij}(\beta_i + \beta_j) + \sigma_{ij}\sqrt{2u_*}(\sqrt{\beta_i} + \sqrt{\beta_j})]\;.
			\end{align*}
		\end{thm}	
		The proof of \Cref{thm:W} proceeds similarly to that of \Cref{thm:X}, albeit with more involved algebra. 
		It is provided in the Supplementary Information.	
		We note that, while it is notationally convenient to leave the final (inside the square root) term unexpanded, the term $\beta_i^2\beta_j^2$ cancels. 
		The entire expression is therefore of polynomial order $-\frac{1}{2}$ in the entries of $\BETA$, and again goes to zero as these entries grow large.  

		Informally, \Cref{thm:W} states that 
		\begin{align}
			\omega_{ij} \approx \frac{f_{ij}(\BETA)}{1 - f_{ij}(\BETA)}\;. \label{eq:w_formula}
		\end{align}
		Recall that $f_{ij}(\BETA) \approx \chi_{ij}$ by \Cref{thm:X}, and that $\chi_{ij} = \eta(w_{ij}\geq 1)$ by definition. 
		Then, \eqref{thm:W} states that $\omega_{ij}$ is approximately equal to the odds that there is at least one edge present between nodes $i$ and $j$. 
	    As we will see, this approximation gives us a method to compute the vector $\BETA$ in terms of the vector $\degree$, thereby obtaining an approximation for the moments of $\W$.
	    As in \Cref{thm:X}, the derived bounds are relatively loose, and  substantially better ones may perhaps be obtained from further analysis. 
	    
	\subsection{Second Moments} \label{subsec:var}
	    Before proceeding, we briefly comment on the $p = 3$ stationarity condition. 
	    From this case on, it becomes quite tedious to control the error terms associated with factoring expectations. 
	    Omitting them, we obtain the approximation 
    	\begin{align*}
    		\E[w_{ij}^2] \approx \omega_{ij}\left(\omega_{ij} + \frac{1}{1 - \chi_{ij}}\right)\;.
    	\end{align*}
    	It follows that
    	\begin{align}
    		\sigma_{ij}^2 = \var{w_{ij}} \approx \frac{\chi_{ij}}{(1 - \chi_{ij})^2} \approx \omega_{ij}(\omega_{ij} + 1)\;. \label{eq:variance}
    	\end{align}
    	Note that, under this approximation, $\sigma_{ij}^2 > \omega_{ij}$ whenever $\chi_{ij} > 0$.
    	It is common to model the entries of the adjacency matrix as Poisson random variables, for which the mean and variance are equal. 
    	The formula \eqref{eq:variance} suggests that this approach will be approximately correct for the uniform model when $\omega_{ij}\ll 1$, but systematically underestimate the variance for larger values.

%% file: tex/algorithms.tex
\section{Estimation of $\BETA$} \label{sec:optimization}
    
	We now possess approximate formulae for the low-order moments of $\W$ in terms of the vector $\BETA$. 
	In practice, we do not observe $\BETA$ and must therefore estimate it from $\degree$. 
	To do so, we impose the degree constraint $\sum_{j}\omega_{ij} = d_i$ and insert the approximation given by \Cref{thm:W}. 
	Eliding the error terms, we obtain 
	\begin{align*} 
		d_i \approx \sum_{j} \frac{f_{ij}(\BETA)}{1 - f_{ij}(\BETA)}\;. 
	\end{align*}	
	We therefore define the function $\mathbf{h}:\R^n_+\rightarrow \R^n$ componentwise as
	\begin{align}
		h_{i}(\BETA) \triangleq \sum_{j} \frac{f_{ij}(\BETA)}{1 - f_{ij}(\BETA)}\label{eq:h}
	\end{align}
	and aim to solve the equation 
	\begin{align}
		\mathbf{h}(\BETA)  = \degree \label{eq:to_solve}
	\end{align}
	for $\BETA$.  
	We define the estimator $\hat{\BETA}$ as the solution of \eqref{eq:to_solve}.
	We then use the estimators $\hat{\chi}_{ij} \triangleq f_{ij}(\hat{\BETA})$ and $\wI_{ij} \triangleq \frac{f_{ij}(\hat{\BETA})}{1 - f_{ij}(\hat{\BETA})}$ supplied by \Cref{thm:X,thm:W} as estimates of the moments of $\mathbf{W}$. 
	In general, $\hat{\BETA} \neq \BETA$, since we have discarded the error terms derived in the previous section.
	We should therefore expect that $\hat{\BETA}$ is a biased estimator of $\BETA$, and that $\WI$ is a biased estimator of $\Eomega$. 
	Experiments, however, will show that these biases are substantially smaller than those of $\WNG$.
	
	To get some intuition on the behavior of \eqref{eq:to_solve}, it is useful to consider two contrasting cases. 
	First, consider the degree sequence $\degree = d\e$.
	In this case, $\mathcal{G}_\degree$ is the set of regular graphs in which all nodes have the same degree $d$. 
	We can find a solution of \eqref{eq:to_solve} analytically. 
	We assume that $\BETA = \beta\e$ for some scalar $\beta$. 
	Then, \eqref{eq:to_solve} reads
	\begin{align*}
		\frac{(n-1)\beta^2}{n\beta - \beta^2} = d\;.
	\end{align*}
	Solving for $\beta$ yields the estimator $\hat{\beta}$: 
	\begin{align*}
		\hat{\beta} =  \frac{d}{1 + n^{-1}(d-1)}\;.
	\end{align*}
	We see that, in a sparse limit in which we let $n \rightarrow \infty$ while $d = o(n)$, $\hat{\beta} \rightarrow d$. 
	This reflects the asymptotic equivalence of uniform and configuration models under large, sparse limits. 
	
	Our second example illustrates a case in which no interpretable solution to \eqref{eq:to_solve} exists. 
	Consider the star graph, which possess $k\geq 2$ leaves (labeled 1 through $k$) and a central node (labeled $k+1$). 
	A single edge connects each leaf to node $k+1$.
	Node $k+1$ has degree $k$, while each leaf has degree $1$. 
	There are no valid edge-swaps, and the corresponding null space $\mathcal{G}_\degree$ therefore contains only one element. 
	We can thus read off the correct expected collapsed degree sequence: 
	$\beta_{k+1} = k$ and $\beta_{j} = 1$ for $1\leq j\leq k$. 
	However, this sequence does not solve \eqref{eq:to_solve}. 
	Indeed, letting $\beta_L$ denote the unknown shared collapsed degree for each leaf and $\beta_C$ the collapsed degree of node $k+1$, we can write \eqref{eq:to_solve} as 
	\begin{align*}
		k &= k\frac{\beta_{C}\beta_L}{2\psi - \beta_C\beta_L} \\ 
		1 &= \frac{\beta_C\beta_L}{2\psi - \beta_C\beta_L} + (k-1)\frac{\beta_L^2}{2\psi  - \beta_L^2}\;.
	\end{align*} 
	The first line requires that $\frac{\beta_C\beta_L}{2\psi - \beta_C\beta_L} = 1$. 
	In conjunction with the second line, this implies that $\beta_L = 0$, which in turn contradicts the first line unless $\beta_C = 0$ as well. 
	We conclude that no solution to \eqref{eq:to_solve} exists which respects the symmetries of the star graph.
	On the other hand, simply adding a second copy of the star graph is sufficient introduce a solution. 
	For example, in the union of two 5-stars, the algorithm we develop below to solve \eqref{eq:to_solve} finds that $\beta_C \approx 3.40$ and $\beta_L \approx 0.93$, with mean-square error below machine precision.
	In light of these examples, the conditions such that $\hat{\BETA}$ exists constitutes an interesting direction for future research. 

	\subsection{Properties of $\hat{\BETA}$}
		While existence remains an open question, it is possible to provide a qualified uniqueness guarantee for \eqref{eq:to_solve}.
		We will also prove several simple results about the ``shape'' of the entries of $\hat{\BETA}$ as functions of the entries of $\degree$. 
		Throughout this section, we assume that $\hat{\BETA}$ is sorted, so that $\hat{\beta}_1\leq \hat{\beta}_2\cdots\leq\hat{\beta}_n$. 
		\begin{dfn}\label{dfn:conditions}
			A vector $\BETA$ is \emph{physical} if $\e \leq \BETA \leq (n-1)\e$ entrywise. 
			A vector $\BETA$ is \emph{well-behaved with parameter $\delta > 0$} if, in addition,  $\beta_n^2 \leq \e^T\BETA - \delta$.  
		\end{dfn}
		The bounds imposed by the physicality condition are in fact obeyed by the true expected collapsed degree vector $\E_\eta[\b]$, provided that $\degree \geq \e$ entrywise.  
		Well-behavedness with parameter $\delta>0$ is sufficient, but not necessary, to ensure that $\hat{\omega}_{ij} = f_{ij}(\hat{\BETA}) (1-f_{ij}(\hat{\BETA}))^{-1} > 0$ for all $i$ and $j$. 
		Let $\mathcal{B}_{\delta}$ denote the set of all physical, well-behaved vectors of (implied) fixed size $n$ with a fixed parameter $\delta > 0$.
		Throughout, we will assume that $\delta$ is ``sufficiently small;'' this will not pose problems due to the inclusion $\mathcal{B}_{\delta'}  \subset \mathcal{B}_\delta$ whenever $\delta' < \delta$. 
		By construction, the function $\mathbf{h}$ defined by \eqref{eq:h} is continuous, and indeed smooth, on $\mathcal{B}_\delta$.  

		We will show that \eqref{eq:to_solve} possesses at most one solution on $\mathcal{B}_\delta$. 
		Let 
		\begin{align}
			\mathcal{L}(\BETA) = \norm{\mathbf{h}(\BETA) - \degree}^2_2	 \label{eq:mse}
		\end{align}
		be the square error associated with approximating $\degree$ by $\mathbf{h}(\BETA)$. 
		Then, the problem 
		\begin{align}
			\min_{\BETA\in \mathcal{B}_\delta} \mathcal{L}(\BETA) \label{eq:opt}
		\end{align}
		achieves its minimum value of $0$ at the solutions of \eqref{eq:to_solve} in $\mathcal{B}_\delta$, provided there are any. 
		We will show that \eqref{eq:opt} possesses at most one such minimum. 

		Our proof relies on an elementary form of the Mountain Pass Theorem \cite{ambrosetti1973dual}, given as Lemma 6.1 in \cite{bisgard2015mountain}. 
		A closely related statement is given as Theorem 5.2 in \cite{jabri2003mountain}.
		
		\begin{dfn}[Palais-Smale Condition, \cite{bisgard2015mountain}] 
			Let $q:\R^n\rightarrow \R$ be a continuously differentiable function. 
			Let $\{\mathbf{a}_n\}$ be a sequence of points in $\R^n$ such that $q(\mathbf{a}_n)$ is bounded and $\norm{\nabla q(\mathbf{a}_n)}\rightarrow 0$. 
			The function $q$ \emph{satisfies the Palais-Smale condition} if any such $\{\mathbf{a}_n\}$ possesses a convergent subsequence. 
		\end{dfn}

		\begin{thm}[Mountain Pass Lemma in $\R^n$, \cite{bisgard2015mountain,ambrosetti1973dual}] \label{thm:mountain_pass}
			Suppose that function $q:\R^n\rightarrow \R$ satisfies the Palais-Smale condition. 
			Suppose further that: 
			\begin{enumerate}
				\item $q(\mathbf{a}_0) = 0$. 
				\item There exists an $r > 0$ and $\alpha > 0$ such that $q(\mathbf{a})\geq \alpha$ for all $\mathbf{a}$ with $\norm{\mathbf{a - a_0}} = r$. 
				\item There exists $\mathbf{a}'$ such that $\norm{\mathbf{a}' - \mathbf{a}_0} > r$ and $q(\mathbf{a}') \leq 0$.  
			\end{enumerate}
			Then, $q$ possesses a critical point $\tilde{\mathbf{a}}$ with $q(\tilde{\mathbf{a}}) \geq \alpha$.  
		\end{thm}
		Our strategy is as follows. 
		We will first show that all critical points of $\mathcal{L}$ are solutions of \eqref{eq:to_solve}. 
		We will then show that all such critical points are, furthermore, isolated local minima of \eqref{eq:opt}. 
		The existence of two such isolated local minima would trigger \Cref{thm:mountain_pass}, implying the existence of an additional critical point with $\mathcal{L}(\BETA) > 0$. 
		Since this is a contradiction, we will conclude that only one such minimum exists.

		Our first step is to lower-bound the eigenvalues of the Jacobian $\mathbf{J}$ matrix of $\mathbf{h}$ at an arbitrary point $\BETA$. 
		This Jacobian may be written 
		\begin{align}
			\mathbf{J} = (\mathbf{S} + \mathbf{D})\left(\mathbf{B}^{-1} - \frac{1}{4\psi}\mathbf{E}\right)\;. \label{eq:deriv} 
		\end{align}
		In this expression, $\mathbf{S}$ is the matrix with entries
		\begin{align*}
			s_{ij} = \begin{cases}
				\frac{f_{ij}(\BETA)}{\left(1 - f_{ij}(\BETA)\right)^2} &\quad i \neq j \\ 
				0  &\quad i = j\;. 
			\end{cases}
		\end{align*}
		We have also defined $\mathbf{D} = \diag \mathbf{S}\e$, and $\mathbf{B} = \diag(\BETA)$. 
		We  note as a point of curiosity that $s_{ij} \approx \var{w_{ij}}$ by \eqref{eq:variance}, although our results here do not depend on this relationship.
		A derivation of \eqref{eq:deriv} is supplied in the Supplementary Information. 
		Let $\lambda_i(\mathbf{M})$ denote the $i$th eigenvalue of the matrix $\mathbf{M}$, sorted in ascending order. 
		Thus, $\lambda_1(\mathbf{M})$ is the smallest eigenvalue of $\mathbf{M}$, and $\lambda_n(\mathbf{M})$ the largest. 

		\begin{lm} \label{lm:jacobian}
			Assume $n\geq 5$. 
			Then, 
			\begin{align}
				\lambda_1(\mathbf{J}) \geq \frac{1}{n(n-1)}\left(1 - \frac{2}{\sqrt{5}}\right) > 0\;.
			\end{align}
			In particular, $\mathbf{J}$ is positive-definite and its eigenvalues are bounded away from zero on $\mathcal{B}_\delta$. 
		\end{lm}
		A proof is given in the Supplementary Information. 
		
		\begin{lm}\label{lm:hessian}
			If $n \geq 5$ and  $\BETA$ is a critical point of $\mathcal{L}$, then 
			\begin{enumerate}
				\item[(a)] $\BETA$ solves \eqref{eq:to_solve}. 
				\item[(b)] The Hessian $\mathbf{H}$ of $\mathcal{L}$ at $\BETA$ is positive-definite. 
			\end{enumerate}
  		\end{lm}
  		\begin{proof}
  			To prove (a), we compute the gradient of $\mathcal{L}$: 
			\begin{align}
				\nabla \mathcal{L}(\BETA) = 2(\mathbf{h}(\BETA) - \degree)^T\mathbf{J}(\BETA)\;. \label{eq:gradient}
			\end{align}
			By \Cref{lm:jacobian}, $\mathbf{J}(\BETA)$ is positive-definite and therefore full-rank on $\mathcal{B}_\delta$.
			It follows that $\nabla\mathcal{L}(\BETA) = 0$ iff $\mathbf{h}(\BETA) = \degree$, or, equivalently, iff $\mathcal{L}(\BETA) = 0$. 

			To prove (b), we calculate the entries of the Hessian. These are 
			\begin{align*}
				\mathbf{H}(\BETA)_{ij} = 2 \sum_{\ell = 1}^n \left[(h_\ell(\BETA) - d_\ell) \frac{\partial^2 h_\ell }{\partial \beta_i \partial \beta_j} + \frac{\partial h_\ell }{\partial \beta_i} \frac{\partial h_\ell}{\partial \beta_j}\right]\;.
			\end{align*} 
			The first term vanishes at critical points. 
			Recognizing the second as an outer product of the rows of $\mathbf{J}$, we can write the Hessian at critical points as 
			\begin{align*}
				\mathbf{H}(\BETA) = 2 \sum_{\ell = 1}^n \mathbf{J}_\ell(\BETA)\mathbf{J}_\ell(\BETA)^T.
			\end{align*}
			Since $\mathbf{J}$ is full rank by \Cref{lm:jacobian}, the sum is full rank and therefore positive-definite. 
			This completes the proof. 
  		\end{proof}
  		We immediately obtain:
  		\begin{cor}
  			If $n \geq 5$, then each critical point of $\mathcal{L}$ is an isolated local minimum, and there are finitely many of them. 
  		\end{cor}
  		For the second clause, we rely on the fact that 
		$\mathcal{B}_\delta$ is closed and bounded. 

		\begin{lm} \label{lm:PS}
			If $n \geq 5$, the restriction of $\mathbf{h}$ to $\mathcal{B}_\delta$ satisfies the Palais-Smale condition.  
		\end{lm}
		\begin{proof}
			Taking norms in \eqref{eq:gradient} and lower-bounding the righthand side, we obtain 
			\begin{align*}
				\norm{\nabla\mathcal{L}(\BETA)}_2 \geq \lambda_1\left(\mathbf{J}(\BETA)\right)\norm{h(\BETA) - \degree}_2\;.
			\end{align*}
			Since $\lambda_1(\mathbf{J}(\BETA))$  is bounded away from zero on $\mathcal{B}_\delta$ by \Cref{lm:jacobian}, the only sequences $\{\BETA_t\}$ in $\mathcal{B}_\delta$ that satisfy $\norm{\nabla \mathcal{L}(\BETA_t)}_2\rightarrow 0$ must also satisfy $\mathbf{h}(\BETA_t)\rightarrow \degree$.
			By \Cref{lm:hessian}, there are finitely many solutions to \eqref{eq:to_solve}, and therefore any such sequence has a finite number of limit points. 
			The sequence $\{\BETA_t\}$ then possesses a subsequence that converge to each of these limit points, which completes the proof.  
		\end{proof}

  		\begin{thm} \label{thm:uniqueness}
  			If $n \geq 5$, there exists at most one $\hat{\BETA}$ in the set $\mathcal{B}_\delta$ such that $\mathbf{h}(\hat{\BETA}) = \degree$. 
  		\end{thm}
  		\begin{proof}
  			Suppose that there were two solutions $\BETA_0$ and $\BETA_1$ in $\mathcal{B}_\delta$. 
  			Since $\mathcal{L}$ satisfies the Palais-Smale condition (\Cref{lm:PS}), we check conditions (1)-(3) of \Cref{thm:mountain_pass} are satisfied. 
  			Condition (1) requires that $\mathcal{L}(\BETA_0) = 0$, which is true by hypothesis. 
  			Condition (2) requires that there exists $r > 0$ and $\alpha > 0$ such that $h(\mathbf{\BETA}) \geq \alpha$ for all $\mathbf{\BETA}$ with $\norm{\BETA - \BETA_0} = r$.
  			This follows from Taylor-expanding $\mathcal{L}$ around $\BETA_0$ and using the positive-definiteness of $\mathbf{H}$. 
  			Applying \Cref{lm:hessian} yields the existence of such an $r$ and $\alpha$, and further implies that $r$ may be taken to be arbitrarily small. 
  			In particular, $r$ may be taken to be smaller than $\norm{\BETA_0 - \BETA_1}$, which in turn supplies condition (3). 
  			Applying \Cref{thm:mountain_pass}, we conclude that there exists a critical point $\tilde{\BETA}$ of $\mathcal{L}$ such that $\mathcal{L}(\tilde{\BETA}) \geq \alpha > 0$. 
  			But this contradicts \Cref{lm:hessian}. 
  			We conclude that at most one solution to \eqref{eq:to_solve} exists in $\mathcal{B}_\delta$, as was to be shown. 
  		\end{proof}

  		Numerical experiments suggest that the solution to \eqref{eq:to_solve}, if it exists, may be unique in the positive orthant $\R^{n}_+$. 
  		If true, this would be a stronger result than that provided by  \Cref{thm:uniqueness}, which requires physicality and well-behavedness. 
  		Extending \Cref{thm:uniqueness} to cover the full nonnegative orthant would be an interesting direction of future work. 

		The following theorem specifies several properties of $\hat{\BETA}$, provided that it exists. 

		\begin{thm} \label{thm:properties}
			Let $n \geq 5$. 
			Suppose that $\hat{\BETA} \in \mathcal{B}_\delta$ solves \eqref{eq:to_solve}. 
			Then, 
			\begin{enumerate}
				\item[(a)] The map $d_i\mapsto \hat{\beta}_i$ is nondecreasing. 
				\item[(b)] Furthermore, $\hat{\beta}_i - \hat{\beta}_j \leq d_i - d_j$\;.
				\item[(c)] Finally, $\hat{\BETA} \leq \degree$ entrywise.
			\end{enumerate}
		\end{thm}
		A proof of this result is given in the Supplementary Information.  
		
	\subsection{Algorithms}
		Having proven some properties of the solutions of \eqref{eq:to_solve}, it remains to develop an algorithm to find these solutions.
		While it is possible to use standard gradient-based methods, this task is complicated by the ill-conditioned Jacobian of $\mathbf{h}$. 
		Ill-conditioning arises from dramatic heterogeneity in the entries of $\mathbf{S}$. 
		For example, in the experiments shown in \Cref{fig:validation} in the next section, the observed and estimated values of $\sigma_{ij}$ span four orders of magnitude, implying that entries of $s_{ij} \approx \sigma_{ij}^2$ span roughly eight. 
		Because of this, methods based on the full Jacobian, such as standard implementations of gradient descent or Newton's method, may require impractically small step-sizes in order to avoid pathological behavior. 

		We instead adopt a coordinate-wise approach. 
		Suppose we have a current estimate $\hat{\BETA}^{(t-1)}$. 
		We obtain an estimate of $\degree$ given by $\hat{\degree}^{(t-1)} = \mathbf{h}(\hat{\BETA}^{(t-1)})$. 
		To update the $i$th coordinate of $\hat{\BETA}^{(t-1)}$, we hold all other $n-1$ coordinates fixed, and define $\hat{\beta}_{i}^{(t)}$ to be the value of $b$ that solves the equation 
		\begin{align}
			h_i\left(\hat{\beta}_1^{(t-1)},\ldots,\hat{\beta}_{i-1}^{(t-1)}, b, \hat{\beta}_{i+1}^{(t-1)}, \ldots,\hat{\beta}_n^{(t-1)}\right) = d_i\;. \label{eq:iteration}
		\end{align}
		We repeat this process for each of the $n$ coordinates, obtaining a fully updated new estimate $\hat{\BETA}^{(t)}$. 
		We iterate this sweep over the coordinates until a desired error function drops below a user-specified tolerance. 
		A standard choice for the error function is the mean-square error $n^{-1}\mathcal{L}(\BETA)$, with $\mathcal{L}$ as in \eqref{eq:mse}. 
		\Cref{alg:opt} formalizes the solution method.
		The call $\text{Solve}_b$ solves a single-variable equation for $b$. 
		In the accompanying code (see ``Software''), we implement $\text{Solve}_b$ with options to use either the \texttt{root\_scalar()} function supplied by \texttt{python}'s \texttt{scipy} package or a bespoke Newton-type method. 

		\begin{algorithm2e}[H]
           \DontPrintSemicolon
            \caption{Computation of $\hat{\beta}$}\label{alg:opt}
            \KwIn{degree sequence $\degree \in \mathbb{Z}_+^n$, initial guess $\Bhat^{(0)} \in \R_+^n$, tolerance $\epsilon$}
            \textbf{Initialization:} $t \gets 0$, $\gamma \gets \infty$\;
            \While{$\gamma^{(t)} > \epsilon$}{
                \For{$i = 1,\ldots,n$}{
                    $\bhat^{(t)}_{i} \gets \text{Solve}_{b} \{h_i(\hat{\beta}_1^{(t-1)},\ldots,\hat{\beta}_{i-1}^{(t-1)}, b, \hat{\beta}_{i+1}^{(t-1)} \ldots,\hat{\beta}_n^{(t-1)}) = d_i\}$ 
                }
                $\gamma^{(t)} \gets n^{-1}\mathcal{L}(\hat{\BETA}^{(t)})$\;
                $t \gets t+1$
            }
            \KwOut{$\Bhat^{(t)}$}
        \end{algorithm2e}

        In order to ensure that this algorithm is well-defined, we will show that the update given by \eqref{eq:iteration} possesses a unique solution under mild conditions. 
   		\begin{lm} \label{lm:update}
   			Assume that $\degree > \mathbf{0}$ and $\BETA^{(t-1)} > \mathbf{0}$ entrywise. 
   			Then, for each $i$, \eqref{eq:iteration} possesses a unique solution in $b$ on the open interval $\left(0, \frac{2\psi^{(t-1)}}{\max_{\ell \neq i}\beta^{(t-1)}_\ell}\right)$. 
   		\end{lm}
   		\begin{remark}
   			The hypotheses of \Cref{lm:update} can be ensured by removing degree-zero nodes from $\degree$ and initializing $\BETA^{(0)} > \mathbf{0}$. 
   		\end{remark}
   		\begin{proof}
   			To prove existence, we note that $h_i$ is a continuous function of $\beta_i$. 
   			We have $h_i(\beta_1,\ldots,0,\ldots \beta_n) = 0$ and
   			\begin{align*}
   				\lim_{\beta\rightarrow \frac{2\psi^{(t-1)}}{\max_{\ell \neq i}\beta_\ell}} h_i(\beta_1,\ldots,\beta,\ldots,\beta_n) = \infty\;.
   			\end{align*}
   			The Intermediate Value Theorem then provides existence. 

   			To show uniqueness, it suffices to check the derivative (cf. \eqref{eq:deriv})
	        \begin{align*}
	         	\frac{\partial h_i(\BETA)}{\partial \beta_i} = \left(\frac{1}{\beta_i} - \frac{1}{2\psi}\right)\sum_{\ell \neq i} \frac{f_{i\ell}(\BETA)}{(1 - f_{i\ell}(\BETA))^2}\;.
	        \end{align*}
   			When $\BETA > 0$, this expression is strictly positive. 
   			The function $h_i$ is therefore strictly increasing on $I$, proving uniqueness.  
   		\end{proof}
   		
   		While we have existence, uniqueness, and convergence guarantees for each coordinate update, we possess no such guarantees for \Cref{alg:opt} as a whole.
   		Additionally, it may be the case that some elements of the sequence $\{\hat{\BETA}^{(t)}\}$ produce estimates $\hat{\omega}^{(t)}$ of the adjacency matrix in which some entries are negative. 
   		However, we have never observed \Cref{alg:opt} to fail to converge to a solution in which all entries of $\hat{\omega}^{(t)}$ are positive. 
   		Additionally, when a solution to \eqref{eq:to_solve} exists in $\mathcal{B}_\delta$ for some $\delta$, we have never observed \Cref{alg:opt} to fail to find this solution. 
   		In practice, an analyst can assess the success of the algorithm by checking that (a) the mean-square error is near zero and that (b) the corresponding estimate $\hat{\Eomega}$ has nonnegative entries.  
   		Both such checks are implemented in the accompanying software. 
   		
   		\IfSubStringInString{\detokenize{main}}{\jobname}{}{
			\bibliographystyle{plain}
			\bibliography{../references}
		}

%% file: tex/results.tex
\section{Experiments} \label{sec:results}
    
    In this section, we describe a sequence of experiments exploring the behavior of \Cref{alg:opt}; the accuracy of the estimator $\hat{\BETA}$; the disparity between $\WNG$ and $\WI$ on empirical networks; and implications for downstream tasks such as modularity maximization.  

	\subsection{Synthetic Data} \label{subsec:results}
\begin{figure}
	        \centering
	        \includegraphics[width=\textwidth]{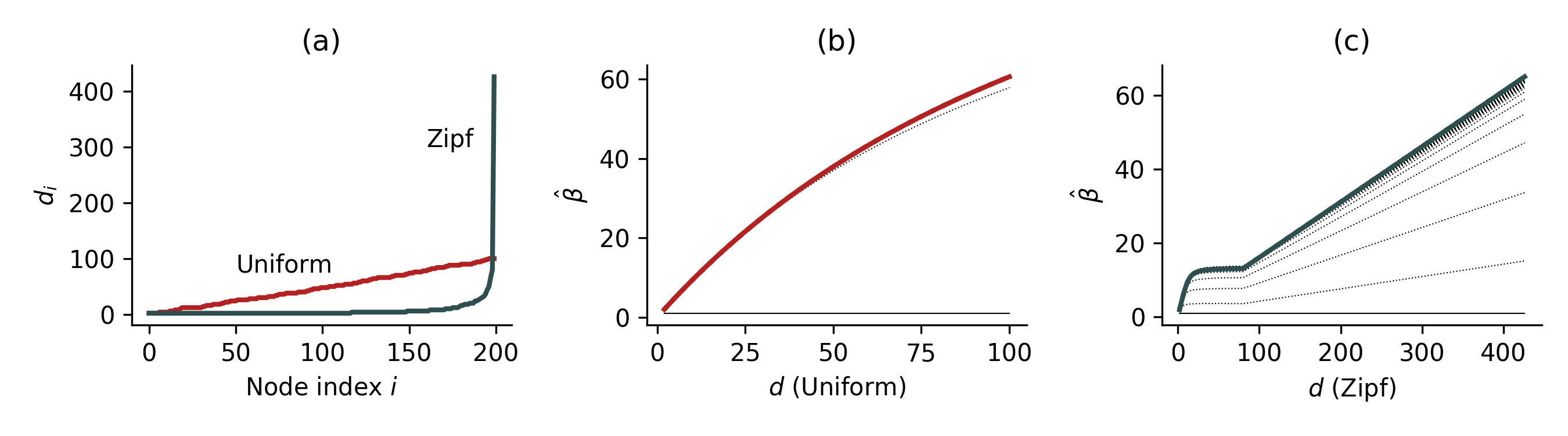}
	        \includegraphics[width=\textwidth]{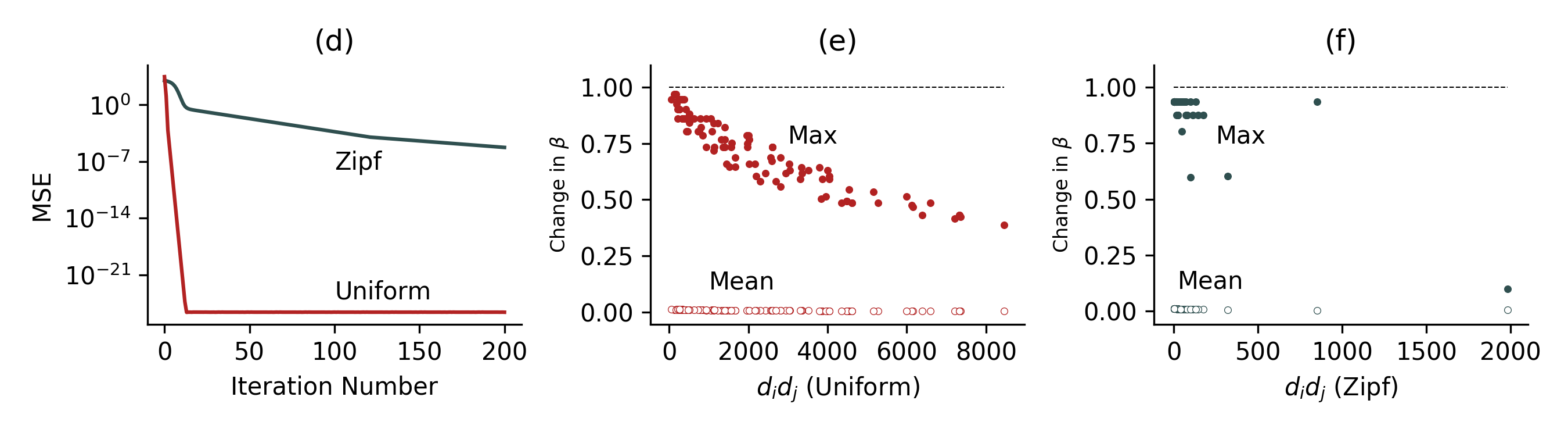}
	        
	        \caption{
	        	(a): The uniform and Zipf degree sequences described in the text. 
	        	(b): Iterates of \Cref{alg:opt} for the uniform degree sequence. 
                The final iterate is highlighted.   
	        	(c): Iterates of \Cref{alg:opt} for the Zipf degree sequence. 
                The final iterate is highlighted. 
	        	(d): Mean-square error (MSE) in \Cref{alg:opt} for the uniform (red) and Zipf (gray) degree distributions as a function of the iteration number.  
	        	(e): Bootstrap estimates of $\norm{\BETA(\degree + \e_i + \e_j) - \BETA}_\infty$ (filled points) and $n^{-1}\norm{\BETA(\degree + \e_i + \e_j) - \BETA}_1$ (empty circles)for the uniform degree sequence. 
                Each point corresponds to a uniformly random choice of distinct indices $i$ and $j$. 
	        	(f): As in (e), for the Zipf degree sequence. 
	        	}
	        \label{fig:convergence}
	    \end{figure}

        To study the convergence behavior of \Cref{alg:opt}, we test it on two synthetic degree sequences.  
        The ``uniform'' sequence consists of $200$ independent copies of $2(u + 1)$, where $u$ is a discrete uniform random variable on the interval $[0, 50]$. 
        We contrast this with a ``Zipf'' sequence $\degree_2$, generated by sampling $200$ copies of $2z$, where $z$ is distributed as a Zipf random variable with parameter $\alpha = 2$. 
        These degree sequences are shown in \Cref{fig:convergence}(a).  
        By design, the uniform sequence is relatively homogeneous in its degrees, while the Zipf sequence possesses a small number of extremely high-degree nodes. 

        We then estimated $\hat{\BETA}$ for each of these degree sequences using \Cref{alg:opt}, initialized with $\BETA^{(0)} = \mathbf{e}$. 
        The estimates for the uniform sequence $\degree_1$ converge rapidly, as shown in panel (b), and after two rounds the iterates cannot be distinguished by eye from the final estimate.
        The final estimate $\hat{\BETA}$ is both physical and well-behaved. 
        By \Cref{thm:uniqueness}, it is the only such solution to \eqref{eq:to_solve}. 
        In contrast, the estimates for the Zipf-distributed sequence $\degree_2$, shown in panel (c), require many rounds to converge. 
        \Cref{fig:convergence}(d) compares the differing convergence rates. 
        The vertical axis gives the mean-square error (MSE) $\frac{1}{n}\norm{\mathbf{h}(\BETA) - \degree}_2$.  
        While the MSE for the uniform degree sequence converges to within machine precision after 14 rounds, the Zipf iterates require over 200 iterations to reach an MSE below $10^{-6}$.
        The resulting estimate $\hat{\BETA}$ is physical but not well-behaved, and \Cref{thm:uniqueness} is therefore insufficient to provide a uniqueness guarantee. 

        \Cref{alg:opt} also allows us to perform some bootstrap-style tests of \Cref{conj:u}.
        Recall that this conjecture asserts that the constant $u(\degree)$, which bounds the effect of perturbations of $\degree$ on $\beta$, is no larger than $1$. 
        The size of $u(\degree)$ in turn influences the tightness of the error bounds derived in \Cref{sec:moments}. 
        \Cref{fig:convergence}(e-f) shows the results of a simple experiment in which we use our estimator $\hat{\BETA}$ as a surrogate for $\BETA$. 
        For each degree sequence, we repeatedly sample $i$ and $j$ from $\binom{[n]}{2}$.
        We then compute $\hat{\BETA}' = \hat{\BETA}(\degree + \mathbf{e}_i + \mathbf{e}_j)$ and compare it to $\hat{\BETA}$. 
        Filled dots show the maximum absolute change, $\norm{\hat{\BETA}' - \hat{\BETA}}_{\infty}$, while empty dots give the mean absolute change $\frac{1}{n}\norm{\hat{\BETA}' - \hat{\BETA}}_1$. 
        Under \Cref{conj:u}, we would expect that $\norm{\hat{\BETA}' - \hat{\BETA}}_{\infty} \leq 1$, which is indeed the case for both degree sequences.
        These results may be viewed as heuristic supports of \Cref{conj:u}.  
        Additionally, $\frac{1}{n}\norm{\hat{\BETA}' - \hat{\BETA}}_1 \ll 1$. 
        This observation suggests the possibility of substantially tightening the error bounds given in \Cref{sec:moments} by controlling the $\ell^1$-norm rather than the $\ell^\infty$ norm, an interesting problem which we leave to future work. 
		
	\subsection{Evaluation on an Empirical Contact Network} \label{sec:experiments}
        Our evaluation data set is a contact network among students in a French secondary school, called \texttt{contact-high-school}  \cite{Mastrandrea-2015-contact,Benson2018}. 
        During data collection, each student wore a proximity sensor. 
        An interaction between two students was logged by their respective sensors when the students were face-to-face and within approximately 1.5m of each other. 
        Edges are time-stamped, although we do not use any temporal information the present experiments. 
        The original data set contains $n = 327$ nodes and $m = 189,928$ distinct interactions.
        
        We first test the accuracy of the estimator $\WI$, using $\WMC$ as a reliable estimate of the true mean $\Eomega$. 
        Because of the scaling issues associated with estimating $\WMC$ on $m \approx 2\times 10^5$ edges, we constructed a data subset based on a temporal threshold $\tau$, chosen to incorporate approximately the last $5\%$ of the original interaction volume.
        The resulting subnetwork has $268$ nodes and $10,026$ edges.
        To estimate the ground-truth moments $\eta_\degree$, we estimated $\WMC$ on the subnetwork from $10^7$ samples at intervals of $10^3$ steps. 
        This computation required approximately one week on a single thread of a modern server. 
        
        \begin{figure}
            \centering
            \includegraphics[width=\textwidth]{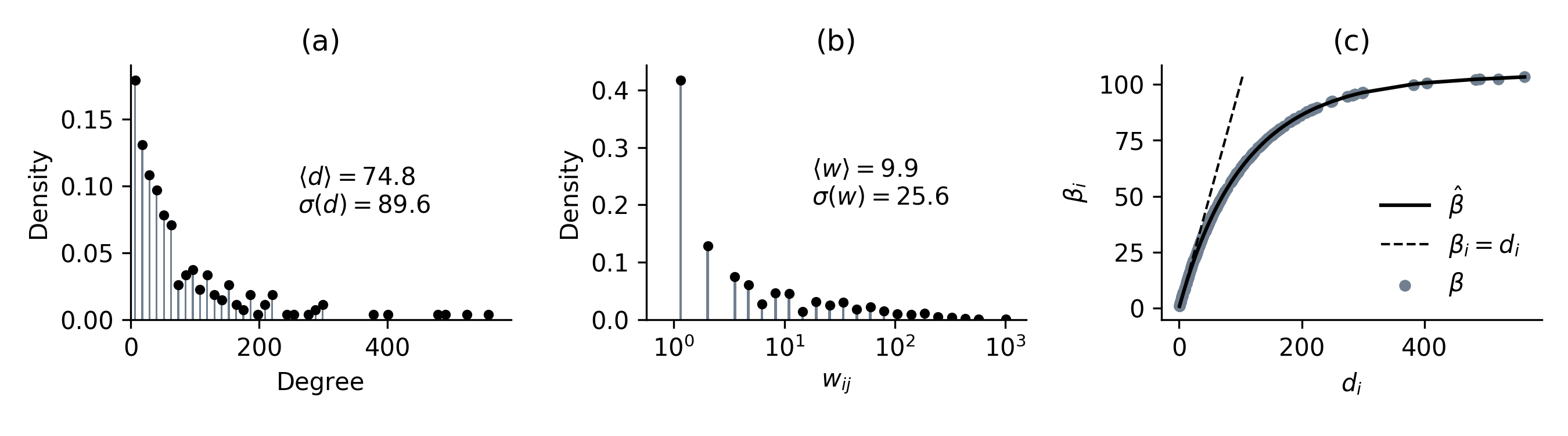}
            \includegraphics[width=\textwidth]{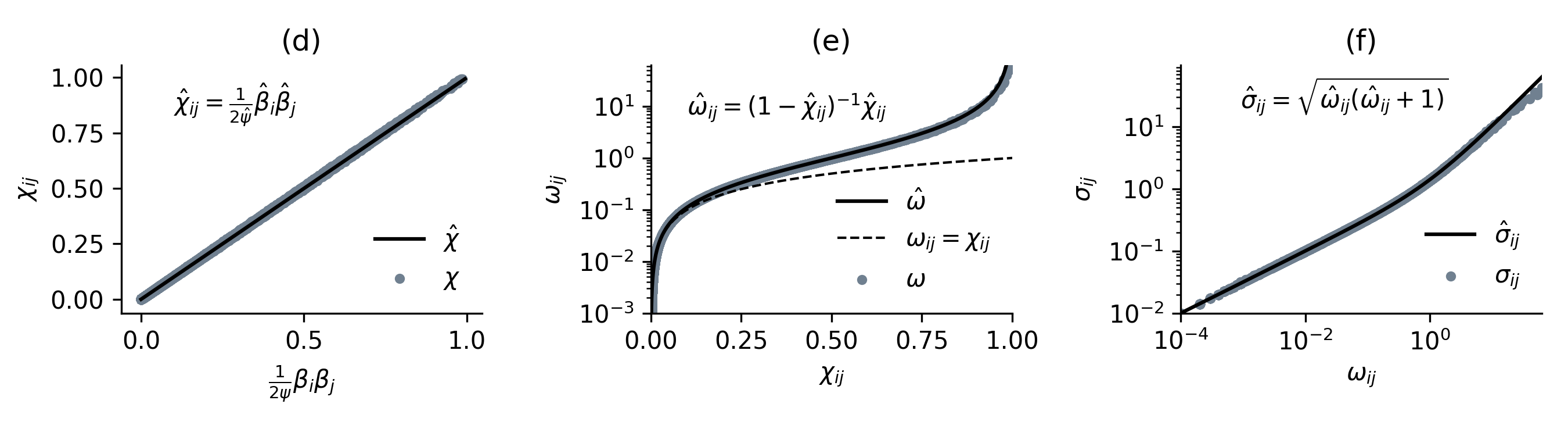}
            
            \caption{
                (a): Degree distribution of the \texttt{contact-high-school} subnetwork.
                The mean degree $\bracket{d}$ and standard deviation of the degree $\sigma(d)$ are shown.  
                (b): Distribution of the entries of $\mathbf{w}$.  
                Note the logarithmic horizontal axis. 
                (c): Collapsed degree sequence $\Bhat$ learned from $\degree$ via \Cref{alg:opt}. 
                Dashes give the line of equality. 
                (d):  Approximation of $\boldsymbol{\chi}$ via \eqref{eq:x_bounds}. 
                (e): Approximation of $\Eomega$ via \eqref{eq:w_formula}. 
                Note the logarithmic vertical axis. 
                Dashes give the line of equality. 
                (f): Approximation of $\sigma_{ij} = \sigma(W_{ij})$ via \eqref{eq:variance}. 
                Note the log-log axis. 
                In (c)-(f), simulated moments (gray dots) are obtained via the Monte Carlo estimator using \Cref{alg:MCMC}; see main text for details. 
                }
            \label{fig:validation}
        \end{figure}
        
        In \Cref{fig:validation}(a)-(b), we show the distributions of degrees and entries of $\mathbf{w}$ for this subnetwork. 
        \Cref{fig:validation}(a) depicts the heterogoenous degree distribution, with standard deviation larger than the average degree.
        While most nodes have small degrees, there are twelve whose degree exceeds $n$.
        \Cref{fig:validation}(b) shows the clumping of edges between pairs of nodes. 
        On average, two students who interact at all interact nearly ten times, but there is substantial deviation around this average. 
        Almost half of all pairs interact just once. 
        In contrast, a small number of pairs interact over 100 times, and one over 1,000. 
        
        In \Cref{fig:validation}(c)-(f), we show the construction of estimators for the moments of $\Eomega$ under the uniform random graph model with the observed degree sequence $\degree$. 
        In \Cref{fig:validation}(c), the solid line shows the estimate $\Bhat$ output by \Cref{alg:opt}, plotted against the degree sequence. 
        Points give the MCMC estimate for $\BETA$. 
        The agreement is almost exact. 
        The estimate $\hat{\BETA}$ is physical and well-behaved.
        \Cref{thm:uniqueness} implies that it is the only physical, well-behaved solution to \eqref{eq:to_solve}. 
        
        In \Cref{fig:validation}(d), we estimate $\hat{\chi}_{ij} \approx f_{ij}(\hat{\BETA}) = \frac{\hat{\beta}_i\hat{\beta}_j}{2\hat{\psi}}$, again finding the agreement to be near exact. 
        In (e), we estimate $\hat{\omega}_{ij}\approx \left(1 - \hat{\chi}_{ij}\right)^{-1}\hat{\chi}_{ij}$. 
        The agreement with data is again excellent, although there is a small amount of visible overestimation of $\omega_{ij}$ when $\chi_{ij}$ is large. 
        Finally, (f) uses \eqref{eq:variance} to compute an estimator $\hat{\sigma}_{ij} = \sqrt{\hat{\omega}_{ij}(\hat{\omega}_{ij} + 1)}$ of $\sigma_{ij}$ the standard deviation of $W_{ij}$. 
        The agreement is strong through roughly $\omega_{ij} \approx 10$, and begins to overestimate $\sigma_{ij}$ for larger values. 
        
        \Cref{fig:validation}(c) and (e) also highlight the relationship of $\WI$ and $\WNG$. 
        The dashed lines in these figures represent two linear approximations that can be made to yield the latter from the former. 
        First, we approximate $\BETA = \degree$ (dashed line, \Cref{fig:validation}(c)). 
        This approximation holds good when $d_i$ is small, since then the number of parallel edges incident to node $i$ should be small -- i.e. $W_{ij} \approx X_{ij}$.  
        Then, we approximate $\Eomega = \boldsymbol{\chi}$ (dashed line, \Cref{fig:validation}(e)). 
        This approximation should hold for small entries of $\Eomega$, since in this case $\chi_{ij}$ is small and $(1-\chi_{ij})^{-1} \approx 1$. 
        As the plots indicate, these approximations are indeed accurate when $d_i$ and $\omega_{ij}$ are small.
        These conditions correspond roughly to the ``large, sparse'' heuristics used frequently in the literature. 
        We can therefore view $\WNG$ as a first-order approximation to $\WI$ near the large, sparse regime. 
        Conversely, we can view $\WI$ as a nonlinear correction to $\WNG$ as we depart from that regime. 
        
\begin{figure}
                \centering
                \includegraphics[width = .6\textwidth]{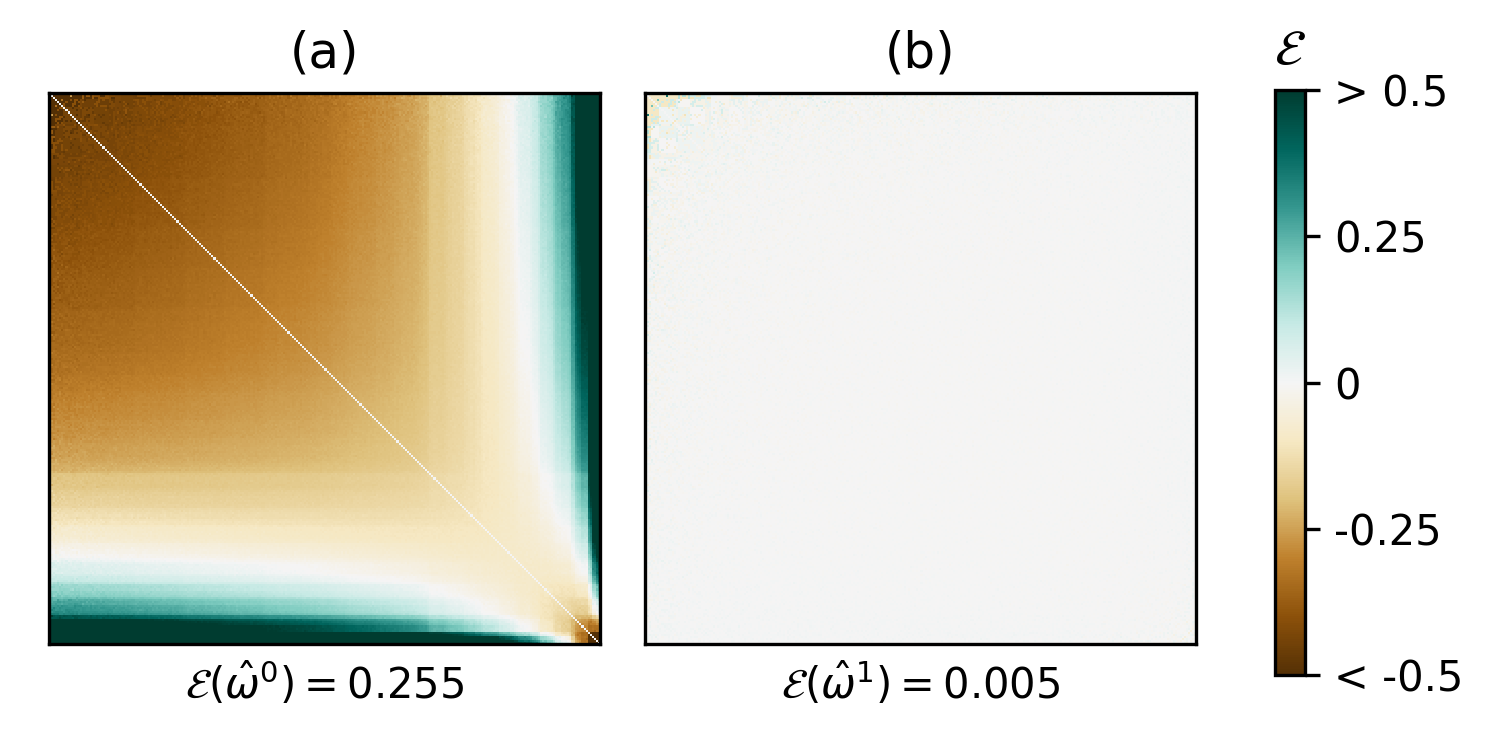}
                \caption{
                    Shading gives the relative error for approximating $\Eomega$ under the uniform model for the \texttt{contact-high-school} subnetwork. 
                    (a): Using the CL estimator $\WNG$. 
                    (b): Using the present estimator $\WI$. 
                    Node degrees in each matrix increase left to right and top to bottom. 
                    The ``ground truth'' is provided by $\WMC$, computed
                    as in \Cref{fig:validation}.
                    }
                \label{fig:predictions}
            \end{figure}

        \Cref{fig:predictions} compares the overall performance of the estimators $\WNG$ and $\WI$. 
        We compute the entrywise relative error $\mathcal{E}_{ij}(\hat{\Eomega}) = (\wMC_{ij})^{-1}(\hat{\omega}_{ij}-\wMC_{ij})$ when approximating $\WMC \approx \Eomega$ with both methods.
        Cells are shaded according to the magnitude and sign of the error.
        The CL estimator in (a) displays systematic bias, underestimating the density of edges between nodes of similar degrees and overestimating the density of edges between nodes with highly disparate degrees. 
        The mean absolute relative error of the Chung-Lu estimate is $\mathcal{E}(\WNG) = \binom{n}{2}^{-1}\sum_{ij}\abs{\mathcal{E}_{ij}(\WNG)} \approx .255$, indicating that a typical entry of $\WNG$ is off by over $25\%$. 
        In contrast, $\WI$ evaluated in (b) has almost no visible bias. 
        Some large residuals are visible for entries $\omega_{ij}$ in which both $d_i$ and $d_j$ are small (top left corner), although it is difficult to evaluate to what extent these residuals reflect error in $\WI$ or in the challenge to $\WMC$ to estimate these edge densities in finite runtime.  
        The mean absolute relative error $\mathcal{E}(\WI)$ is roughly $0.5\%$, an improvement over $\WNG$ of a full order and a half of magnitude. 
            

            

    \subsection{Application: Modularity Maximization in Dense Contact Networks}
    
        Let $\ell:N \rightarrow \mathcal{L}$ be a function that assigns to node $i$ a label $\ell_i \in \mathcal{L}$. 
        The \emph{modularity} of the partition $\ell$ with respect to matrix $\mathbf{w}$ and null model $\rho$ is given by 
        \begin{align}
            Q(\ell;\rho) = \frac{1}{2m}\Sum_{ij}\left[w_{ij} - \E_{\rho}[W_{ij}]\right]\mathbbm{1}(\ell_i, \ell_j)\;. \label{eq:modularity}
        \end{align}
        The normalization by $2m$ ensures that $-1 \leq Q(\ell;\rho) \leq 1$.
        Intuitively, $Q(\ell;\rho)$ is high when nodes that are more densely connected than expected by chance (under the specified null) are grouped together.  
        Maximizing this quantity with respect to $\ell$ may therefore be reasonably expected to identify modular (``community'') structure in the network \cite{Newman2006, Newman2003}.
        Exact modularity maximization is NP-hard \cite{brandes2007finding} and subject to theoretical limitations in networks with modules of heterogeneous sizes \cite{Fortunato}. 
        Despite this, it remains one of the most popular methods for practical community detection at scale \cite{Blondel2008}.

        In most implementations, $\rho$ is not explicitly specified -- rather, the expectation $\E_\rho[W_{ij}]$ is ``hard-coded'' as equal to $\wNG_{ij}$. 
        From a statistical perspective, this reflects an implicit choice of $\rho$ as the Chung-Lu model \cite{Chung2002}, which preserves expected degrees and indeed possesses the given first moment.\footnote{We note that alternative justifications of the use of $\WNG$ exist, including connections to the stability of Markov chains \cite{Delvenne2008} and to stochastic block models \cite{Newman2016a}.} 
        Modifications are possible; the best known is perhaps the resolution adjustment that replaces $\WNG$ with $\gamma \WNG$ for some $\gamma >0$ \cite{reichardt2006statistical}.   
        Other adjustments may incorporate spatial structure \cite{expert2011uncovering} or adjust for the inclusion of self-loops in the null space \cite{cafieri2010loops}. 
        When we wish to perform modularity maximization against a null that deterministically preserves degree sequences, however, $\WNG$ is at best an approximation. 
        We expect this approximation to perform adequately for the configuration model (cf. \Cref{thm:stub}), and very poorly for the uniform model (previous subsection). 
        The Monte Carlo estimate $\WMC$ can be used for very small data sets, but rapidly becomes computationally infeasible for larger ones. 
        In these cases, we can use the present estimator $\WI$ instead. 

        Recent work has highlighted the importance of studying the \emph{modularity landscape}, especially the set of local maxima of $Q$, rather than restricting attention to a single partition.
        One reason for this is the phenomenon of \emph{degeneracy} -- in many practical contexts, a given network will possess many distinct local maxima with modularity comparable to the global maximum \cite{good2010performance}. 
        A second reason is model-specification. 
        As shown in \cite{Newman2016a}, maximization of $Q$ is related to maximum-likelihood inference in a planted-partition stochastic blockmodel. 
        When the planted-partition model is unrealistic as a generative model for the data, modularity-maximization amounts to inference in a mis-specified model. 
        Degeneracy is a common symptom of this problem, but observing it requires locally optimizing $Q$ multiple times. 
        For these reasons among others, ensemble-based methods that implicitly average over local optima, such as those of \cite{zhang2014scalable}, may be preferable.
        With these considerations in mind, our aim in this section is not to show that the use of $\WI$ is strictly superior for this task when compared to $\WNG$ for either one-shot or ensemble-based modularity maximization. 
        Rather, we will argue that the corresponding modularity landscapes are significantly different on data sets of practical interest, and that it is therefore methodologically unsafe to interchange these estimators without carefully scrutinizing the results. 
        
        Our first data set for this experiment is the full \texttt{contact-high-school} network, consisting of $n = 327$ nodes and $m = 189,928$ edges as described in the previous subsection. 
        The computation of $\WMC$ is indeed infeasible for a graph this dense, and we therefore use $\WI$ as an estimate.
        This setting highlights the utility of $\WI$, since otherwise we would have no practical way to compute the uniform expectation. 
        
        \begin{figure}
            \centering
            \includegraphics[width=\textwidth]{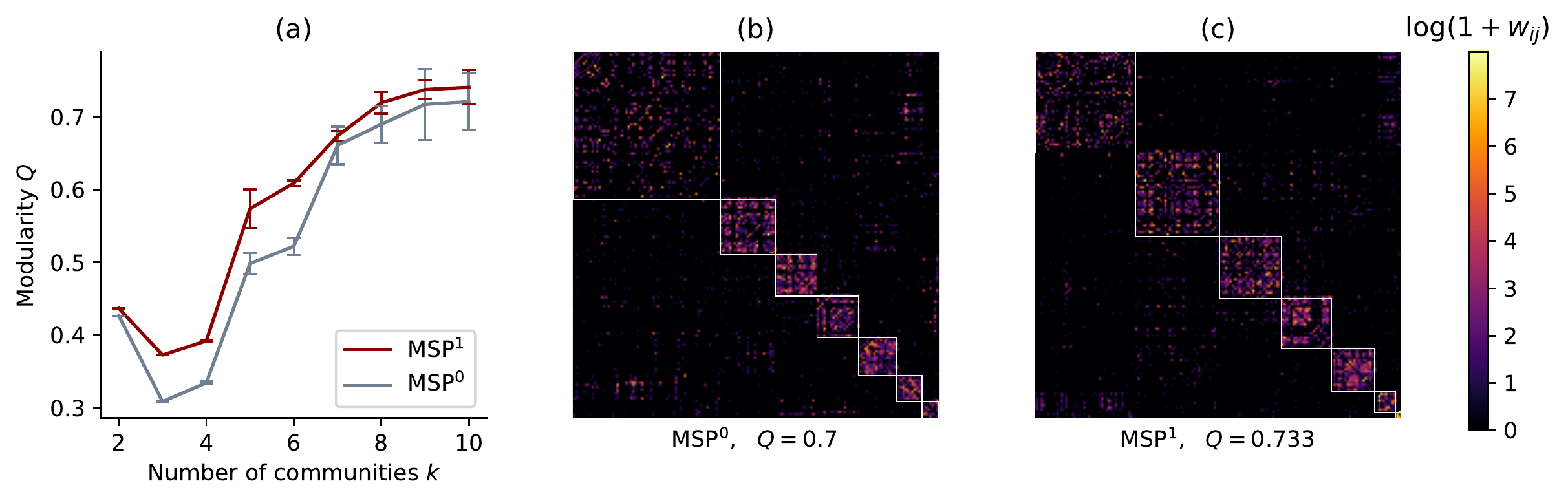}
            \caption{(a): Performance of MSP on the full \texttt{contact-high-school} network using the CL modularity matrix $\mathbf{M}^0$ and the approximate uniform modularity matrix $\mathbf{M}^1$, over 100 batches of 50 repetitions each. 
            Solid lines give the average modularity, and error bars give two standard deviations from the mean. 
            (b): Example partition using $\mathbf{M}^0$. 
            (c): Example partition using $\mathbf{M}^1$. 
            To generate (b) and (c), the best partition of 500 runs was chosen for each algorithm variant. 
            Each run was initialized with $k = 8$; in the best partition, however, only $7$ labels are actually used.  
            Colors are shown on a log scale. 
            }
            \label{fig:modularity}
        \end{figure}

        To approximately maximize \eqref{eq:modularity}, we employ the multiway spectral partitioning (MSP) algorithm of \cite{zhang2015multiway}, which generalizes the spectral bipartitioning algorithm of \cite{Newman2006}. 
        While greedy methods often enjoy superior performance \cite{Blondel2008}, spectral methods have the advantage of depending strongly on the structure of the observed graph and the null model employed, and are relatively insensitive to choices made during the runtime of the algorithm. 
        Spectral methods are therefore ideal for highlighting differences in the modularity landscapes induced by alternative null models. 
        The algorithm requires the analyst to specify a null model and a desired number of communities $k$. 
        The core of the approach is to use a low-rank approximation of the \emph{modularity matrix}, $\mathbf{M} = \w - \E_{\rho}[\W]$.
        This approximation induces a map from the vertices of $G$ to a low-dimensional vector space. 
        Vectors in this space are clustered according to their relative angles using a procedure reminiscent of $k$-means to produce the community assignment. 
        Because the clustering algorithm involves a stochastic starting condition, it is useful to run the algorithm multiple times and choose the highest modularity partition from among the repetitions.
        We refer the reader to \cite{zhang2015multiway} for details, and to the code accompanying this paper for an implementation of MSP for arbitrary modularity matrices (see ``Software'').
        
        We ran this algorithm using both the CL modularity matrix $\mathbf{M}^{0} = \w - \WNG$ and the approximate uniform modularity matrix $\mathbf{M}^{1} = \w - \WI$. 
        We refer to these two algorithmic variants as $\mathrm{MSP}^0$ and $\mathrm{MSP}^1$, respectively. 
        Since $\WNG$ and $\WI$ produce very different null matrices, the modularity matrices $\mathbf{M}^0$ and $\mathbf{M}^1$ are themselves very different -- the mean absolute relative error of using the latter to estimate the former is approximately $32\%$. 
        We would therefore expect $\mathrm{MSP}^0$ and $\mathrm{MSP}^1$ to behave very differently in this task. 
        We allowed the number of communities $k$ to vary between $2$ and $10$. 
        For each value of $k$, we ran $\mathrm{MSP}^0$ and $\mathrm{MSP}^1$ in 100 batches of 50 repetitions. 
        From each batch of 50, the highest-modularity partition was chosen, resulting in 100 partitions per value of $k$.
        \Cref{fig:modularity}(a) shows that $\mathrm{MSP}^1$ tends to find higher modularity partitions than  $\mathrm{MSP}^0$ on this data set.  
        The difference is especially large when $k$ is small, but a substantial difference between the means is noticeable even for larger values.
        While partitions under $\mathbf{M}^0$ exist that are comparable to those under $\mathbf{M}^1$, it appears to be more difficult for $\mathrm{MSP}^0$ to find them. 
        Panels (b) and (c) shed some light on the differing behavior of the two algorithms. 
        Partitions under $\mathrm{MSP}^0$ tends to display a larger, less cohesive community ((b), top left) alongside smaller, more tightly interconnected ones. 
        Partitions under $\mathrm{MSP}^1$ (c) tend to display communities that are slightly more uniform in size. 
        
        It is reasonable to object that  modularity values under $\mathrm{MSP}^1$ and $\mathrm{MSP}^0$ should not be compared, since these objectives are defined with respect to differing null matrices.
        In this specific case, the objection is not borne out numerically, however -- ``cross-evaluating'' the partitions on the opposite matrices changes the modularities only minimally. 
        Evaluating the $\mathrm{MSP}^0$ partition in \Cref{fig:modularity}(b) on the modularity matrix $\mathbf{M}^1$ gives $Q = 0.699$, while evaluating the $\mathrm{MSP}^1$ partition on $\mathbf{M}^0$ yields $Q = 0.731$. 
        On this data set, $\mathrm{MSP}^1$ searches the energy landscape of $\mathrm{MSP}^0$ more efficiently than does $\mathrm{MSP}^0$ itself. 
        
\begin{figure}
            \centering
            \includegraphics[width=\textwidth]{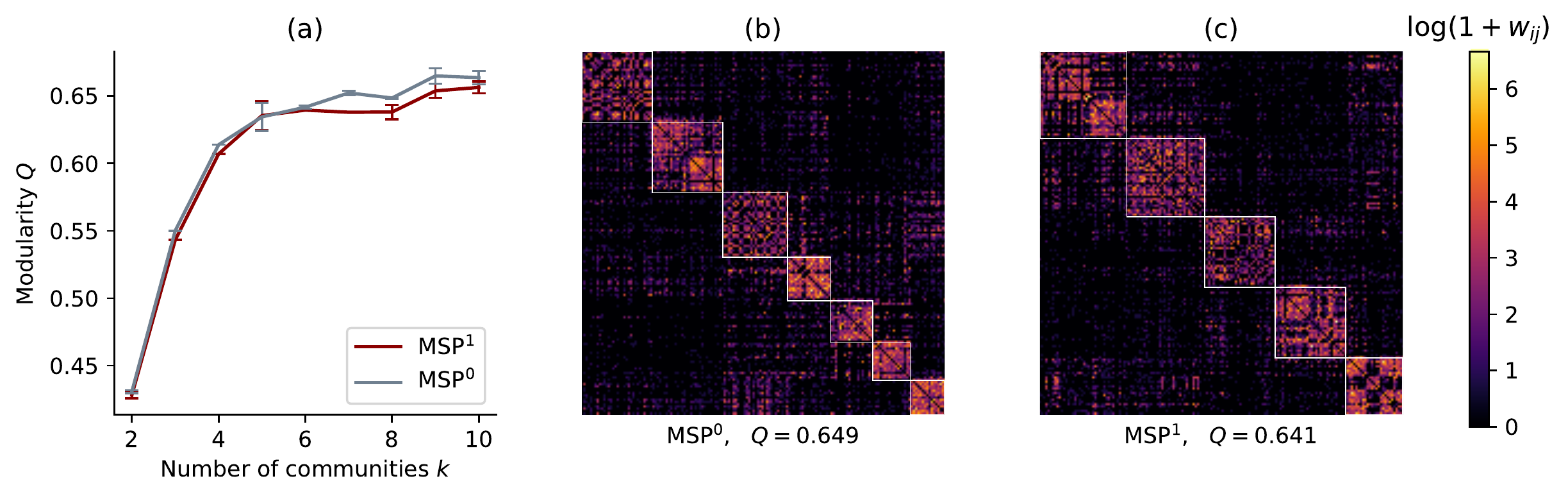}
            \caption{This figure is in all methodological details identical to \Cref{fig:modularity}, using the study data set \texttt{contact-primary-school} \cite{Stehle2011, Benson2018}.
            In (b)-(c), both algorithms were initialized with $k = 8$. 
            }
            \label{fig:modularity_primary}
        \end{figure}

        It should be noted that this behavior is data-set dependent. 
        The opposite case occurs in the \texttt{contact-primary-school} network \cite{Stehl-2011-contact, Benson2018}, which used similar sensors to construct an interaction network among students in a French primary school. 
        On this data, $\mathrm{MSP}^0$ and $\mathrm{MSP}^1$ perform similarly for $k \leq 6$ communities (\Cref{fig:modularity_primary}), with the former consistently outperforming  the latter for $k \geq 7$.  
        The illustrative partitions in panels (b) and (c) suggest $\mathrm{MSP}^1$ tends to prefer partitions with fewer communities. 
        Whereas $\mathrm{MSP}^0$ chooses a partition with $7$ communities, $\mathrm{MSP}^1$ chooses one with just $5$ (both having been initialized at $k = 8$). 
        These illustrations emphasize that $\mathrm{MSP}^1$ and $\mathrm{MSP}^0$ explore different modularity landscapes; that the relative advantages of each algorithm depend on the data; and that the landscape for $\mathrm{MSP}^1$ can be tractably computed under the methodology we have introduced here. 
        
        \IfSubStringInString{\detokenize{main}}{\jobname}{}{
            \bibliographystyle{plain}
            \bibliography{../references}
        }

%% file: tex/discussion.tex
\section{Discussion} \label{sec:discussion}
    
    Much existing network theory is explicitly designed for large, sparse data.
    However, many networks of interest are sufficiently dense to diverge significantly from the predictions of large, sparse theory.
    We have highlighted this phenomenon in the context of dense multigraphs, with a focus on estimating the expected adjacency matrix $\Eomega$ of a random multigraph with specified degree sequence. 
    We have shown that, rather than falling back to computationally expensive MCMC, we can construct an accurate estimator $\WI$ using an indirect, dynamical approach. 
    Use of this estimator can in turn have significant impact on the results of downstream data analyses. 
    
    There are several directions for future work on the moments of uniform random graphs with fixed degree sequences. 
    As previously noted, the error bounds on $\hat{\boldsymbol{\chi}}$ and $\WI$ derived in \Cref{sec:moments} appear quite loose when  when compared against the empirical results in \Cref{fig:convergence,fig:validation}. 
    The derivation of tighter error bounds would be helpful for researchers seeking practical accuracy guarantees. 
    Progress on this front appears to be hindered by the complex combinatorial structure of the space $\mathcal{G}_\degree$; 
    however, carefully chosen assumptions or approximations may allow headway.
    An additional avenue of exploration concerns the impact of the choice between $\WI$ and $\WNG$ on downstream analyses. 
    We have seen that the choice of null expectation can substantially change the performance of $\mathrm{MSP}$, and that the direction of this effect depends on the data set. 
    A better understanding of the properties of the data or algorithm that make certain estimators highlight better solutions would be most welcome. 

    We focused our attention on the derivation of an estimator for $\Eomega$. 
    It may also be possible to derive expressions for higher moments using the same methodology.
    Such moments would approximate expected densities of various motifs under the uniform model.
    Examples of interest may include wedge densities $\E[w_{ij}w_{jk}]$ and  triangle densities $\E[w_{ij}w_{jk}w_{ik}]$. 
    Parsing the stationarity conditions for these more complicated moments may be correspondingly more difficult. 
    An alternative would be to construct mean-field estimates by computing the relevant statistics on $\WI$ itself. 
    An evaluation of the accuracy of this approach would potentially replace the need for computationally intensive MCMC sampling to estimate these quantities. 

    Finally, it may also be of interest to develop similar theory for uniform distributions over related spaces of graphs.
    For example, it would be possible to consider a uniform model including self-loops. 
    The associated analysis would be nontrivial due to required modifications in the MCMC sampling procedure (see \cite{Fosdick2018}), but one might reasonably hope to obtain parallel results. 
    Directed graphs offer another important direction of generalization. 
    It would be natural to define a uniform distribution over spaces of directed multigraphs with fixed in-degree and out-degree sequences. 
    In this case, one might expect analysis to produce expressions for the moments of this distribution in terms of two collapsed degree sequences, corresponding to in- and out-degrees. 
    These and other generalizations offer promising avenues of future work.